\documentclass{aa} 

\usepackage[colorlinks=true, linkcolor=blue, citecolor=blue, urlcolor=blue]{hyperref}

\usepackage[varg]{txfonts}
\usepackage{graphicx}

\usepackage{subcaption}
\usepackage{comment}
\usepackage{booktabs}
\usepackage{bm}
\usepackage{xcolor}
\usepackage{ulem}

\begin{document}

\title{Interpreting the gamma-ray emission from Westerlund 1: \\ hadronic or leptonic?}
\titlerunning{}
\authorrunning{}

\author{Ben Li\inst{1,2,3} \and Pasquale Blasi\inst{1,2} \and Elena Amato\inst{3}}

\institute{Gran Sasso Science Institute (GSSI), Viale Francesco Crispi 7, 67100 L’Aquila, Italy \and INFN-Laboratori Nazionali del Gran Sasso (LNGS), via G. Acitelli 22, 67100 Assergi (AQ), Italy \and INAF-Osservatorio Astrofisico di Arcetri, Largo E. Fermi 5, 50125 Firenze, Italy}

\date{\today}

\abstract
{High-energy gamma-ray emission has been detected from a number of massive star clusters, indicating that these sources are sites of cosmic-ray acceleration. However, it is still unclear whether the gamma-ray emission is of hadronic or leptonic origin.} 
{In this work, we present a comprehensive study of the gamma-ray emission from Westerlund 1 (Wd 1), under the assumption that high-energy particles (both protons and electrons) are accelerated at the cluster wind termination shock. We investigate both the hadronic and leptonic scenarios by fitting simultaneously the spectrum and morphology of the gamma-ray emission.} 
{We numerically solve the transport equations for non-thermal protons and electrons for a given $K_{\mathrm{ep}}$ (electron-to-proton ratio) and derive their energy and spatial distributions. Gamma rays are produced either by hadronic pp interactions of protons with the target gas or by inverse Compton scattering of relativistic electrons (both primary and secondary) with the ambient photon fields. We also calculate the synchrotron emission from these electrons. The predicted spectral energy distributions (SEDs) are compared with the gamma-ray measurements by Fermi-LAT and HESS and with the X-ray upper limit derived by eROSITA. We also compute the neutrino flux from charged pion decay and estimate its detectability with KM3NeT.}
{For reasonable model parameters, we find that in the case of Bohm diffusion, both the leptonic and hadronic scenarios can account for the observed multi-wavelength spectrum and gamma-ray morphology. However, in the case of Kraichnan diffusion, better motivated in star clusters, only the hadronic scenario can explain the observations, while the leptonic scenario fails to reproduce the highest-energy gamma-ray data points.
For the leptonic scenario, the magnetic field immediately downstream of the shock can be constrained to be $\lesssim$ a few $\mu$G thanks to the eROSITA upper limit. On the other hand, in order to explain the gamma-ray observations with hadronic emission alone, the gas density inside the bubble should be $\gtrsim 1.2 \, \mathrm{cm}^{-3}$. A $\sim3.8\sigma$ detection of the corresponding neutrino emission is expected after 10 years of observation with KM3NeT.}
{}

\keywords{acceleration of particles -- diffusion -- ISM: cosmic rays -- gamma rays: ISM -- Galaxy: open clusters and associations: individual: Westerlund 1}

\maketitle
\nolinenumbers

\section{Introduction}

Cosmic rays (CRs) with energies below the "knee" ($\sim 3~\mathrm{PeV}$) are commonly believed to originate from Galactic sources \citep{Hillas_1984,Gaisser_2016}, with those capable of accelerating particles to PeV energies being referred to as PeVatrons. In the standard paradigm, CRs are accelerated by diffusive shock acceleration (DSA) at the forward shock of supernova remnant (SNR) \citep{Bell_1978,Blandford_1978,Drury_1983,Malkov_2001}. However, this scenario can only account for particles with energies below tens of TeV for typical SNRs \citep{Lagage_1983}, while the origin for PeV particles still remains unclear. In the past decades, several magnetic-field amplification mechanisms were proposed to accelerate particles to higher energies, including
the non-resonant streaming instability (\citealt{Bell_2004}; see also \citealt{Amato_2009} for a kinetic treatment) and acoustic instability/kinematic dynamo \cite[]{DruryFalle1986,Jones2009,DruryDownes2012,Capanema2026}.
Although these mechanisms can significantly increase the maximum particle energy, accelerating particles to PeV energies remains challenging for typical SNRs and generally requires rather extreme physical conditions \citep{Bell_2013,Schure_2013,Cristofari2020,Cristofari2021}. Therefore, it is necessary to search for other more efficient CR accelerators, among which pulsar wind nebulae (PWNe), microquasars (MQs) and star clusters (SCs) have attracted increasing attention.

Star clusters are gravitationally bound systems composed of numerous stars, many of which are massive stars ($\gtrsim 8 M_\odot$), including OB-type stars and Wolf-Rayet (WR) stars \citep{PortegiesZwart_2010}. Massive stars lose mass by launching stellar winds into the intracluster medium, forming stellar wind termination shocks (WTSs), where DSA can take place \citep{Klepach_2000}. For compact star clusters, the cluster core occupies a relatively small volume (with a half-mass radius of a few parsecs), in which case the stellar winds can collide and interact with each other, resulting in the formation of a powerful collective wind. The collective wind terminates by interacting with the circumcluster medium and inflates a hot bubble with size of tens of parsec around the cluster \citep{Weaver_1977}. The resulting collective wind termination shock serves as another site for efficient particle acceleration in star clusters \citep{Morlino_2021}. On the other hand, shock accelerated particles propagating in the bubble can scatter with magnetic turbulence and gain energy by second-order Fermi acceleration \citep{Bykov&Toptygin_2001}. 
As a consequence, star clusters host various acceleration sites, making them promising sources for studies of Galactic CRs \citep{Westerlund2022}.

Recently, a number of star clusters have been detected as high-energy gamma-ray sources, further supporting them as CR accelerators. In particular, LHAASO has detected gamma-ray photons with energies up to $\sim 2~\mathrm{PeV}$ from the direction of the Cygnus OB2 association \citep{LHAASO_2024}. \cite{Li_2026} modeled both the gamma-ray spectrum and morphology with hadronic scenario in which gamma rays are produced by proton-proton interactions. They investigated different injection and acceleration models for the Cygnus OB2 association and also introduced the contribution from Galactic CRs. Their results show that extreme conditions are needed to explain the observations in all the cases considered (e.g., suppressed diffusion), and that PeV gamma-ray photons are mainly from Galactic CRs \citep{Li_2026}, although another, more distant source, the microquasar Cygnus X-3, may also provide a non-negligible contribution \citep{LHAASO_cygx3, Shi_cygx3}. On the other hand, high-energy gamma rays can also be produced by inverse Compton (IC) scattering of relativistic electrons. \cite{Harer_2025} fitted the gamma-ray spectrum of Cygnus OB2 using their hybrid model (i.e., both leptonic and hadronic); however, they did not show the gamma-ray radial profile. Therefore, whether the gamma-ray emission from the Cygnus association is leptonic or hadronic still remains under debate.

At the same time, HESS published its updated gamma-ray observation of HESS J1646--458 \citep{Westerlund2022}, which is spatially coincident with Westerlund 1 (Wd 1), one the most massive young star clusters in the Galaxy. HESS identified a shell-like structure around Wd 1, which possibly represents the location of the collective wind termination shock. They also showed that there is no clear correlation between the gamma-ray emission and gas clouds inferred from HI and CO observations \citep{Westerlund2022}. This indicates that the gamma-ray emission favors a leptonic origin. \cite{Harer_2023} compared the hadronic and leptonic scenarios and favored the latter; however, the gamma-ray morphology was not investigated in detail. \cite{Lemoine-Goumard_2025NatCo..1610820L} reanalyzed the Fermi-LAT data toward Wd 1 and identified two gamma-ray components. They proposed that the emission is due to a cosmic-ray loaded nascent outflow driven by Wd 1 and also interpreted the gamma-ray emission as leptonic. However, a more recent work reproduced both the gamma-ray spectrum and morphology assuming a hadronic scenario. The required gas density inferred from their analysis is $\gtrsim 1.1 \, \mathrm{cm}^{-3}$ \citep{Shi_2026JHEAp..5100560S}. Furthermore, \cite{Sano_2026ApJ..1004...12S} reanalyzed the gas distribution around Wd 1 and identified two cloud components with different $v_\mathrm{LSR}$\footnote{Velocity with respect to the local standard of rest (LSR).} which they explained as cloud-cloud collision. They also interpreted the "HI-dip" corresponding to the peak of the southeastern gamma-ray shell in the HESS map as the HI self-absorption. The correlation between the total ISM proton column density and the gamma-ray flux further supports the hadronic explanation \citep{Sano_2026ApJ..1004...12S}. Therefore, the gas distribution toward Wd 1 appears to be more complex than indicated by previous analyses. As a result, the hadronic scenario cannot be ruled out and requires further investigation.

Existing studies either focus on the leptonic or hadronic interpretation of the gamma-ray spectrum of Wd 1, or adopt simplified descriptions of particle acceleration and transport. A self-consistent comparison of the two scenarios using updated multi-wavelength observations within a unified particle acceleration and transport framework is still lacking.
In this work, we present a comprehensive study of the multi-wavelength spectrum and gamma-ray morphology of Wd 1, assuming particle acceleration at the collective wind termination shock, for both cases of leptonic and hadronic scenarios. The gas distribution toward Wd 1 is reanalyzed, including three different hydrogen gases: atomic, molecular and ionized hydrogen. We discuss two different diffusion models (i.e., Bohm and Kraichnan diffusion) in both scenarios. The corresponding neutrino flux in the hadronic scenario is predicted, and we assess its future detectability with KM3NeT.

This paper is organized as follows. In Section \ref{sec:model}, we introduce the model applied to describe particle acceleration at the collective wind termination shock of star cluster. In Section \ref{sec:env}, We describe the cluster environment of Wd 1. The leptonic and hadronic scenarios are presented separately in Section \ref{sec:lep} and Section \ref{sec:had}. Discussion and conclusions are presented in Section \ref{sec:discuss}.

\section{Model description}
\label{sec:model}

For a compact young massive star cluster (YMSC), e.g., Wd 1, the winds of the individual stars can form a collective wind that interacts with its surrounding medium and forms a rarefied hot bubble with extension of tens of pc. Assuming spherical symmetry, the size of the bubble can be estimated following the method given in Appendix \ref{app:evolution} \citep{Castor_1975, Weaver_1977, Gupta_2018,Morlino_2021}:
\begin{equation}
    R_{\mathrm{b}}(t) \simeq 139 ~ \xi_{\mathrm{b}}^{1/5} ~ n_{10}^{-1/5} ~ \dot{M}_{-4}^{1/5} ~ v_8^{2/5} ~ t_{10}^{3/5} ~ \mathrm{pc},
    \label{eq:bubble radius}
\end{equation}
where $n_{10}=n_0/(10 ~ \mathrm{cm}^{-3})$ is the number density of the ambient gas, $\dot{M}_{-4}=\dot{M}/(10^{-4} ~ M_{\odot} ~ \mathrm{yr}^{-1})$ is the total mass loss rate of a star cluster and $v_8=v_\mathrm{w}/(1000 ~ \mathrm{km} ~ \mathrm{s}^{-1})$ is the constant plasma velocity upstream of the shock. The quantity $t_{10}=T_\mathrm{age}/(10 ~ \mathrm{Myr})$ is  the dynamical time of the system. $\xi_{\mathrm{b}} \approx 0.22$ is a numerical factor that accounts for cooling of the hot gas inside the bubble \citep{Vieu+2022}. For Wd 1, we use $\dot{M} = 5 \times 10^{-4} ~ M_{\odot} ~ \mathrm{yr}^{-1}$, $v_\mathrm{w} = 2500 ~ \mathrm{km} ~ \mathrm{s}^{-1}$, and $T_\mathrm{age} = 4 ~ \mathrm{Myr}$ throughout this paper \citep{Harer_2023}. The gas density in the cluster is highly uncertain and inhomogeneous and can be as high as $n_0 \sim 190 \, \mathrm{cm}^{-3}$ \citep{Westerlund2022}.
Thus, here we regard $n_0$ as a free parameter, which can be determined by fitting the gamma-ray morphology ($n_0 \sim 30-50 \, \mathrm{cm}^{-3}$, see Table \ref{tab:para}). 

The bulk energy of the supersonic cluster wind is dissipated at a termination shock, whose radius can be estimated by equating the ram pressure of the upstream wind with the pressure inside the bubble:
\begin{equation}
    R_{\mathrm{s}}(t) \simeq 24.3 ~ \xi_{\mathrm{b}}^{-1/5} ~ n_{10}^{-3/10} ~ \dot{M}_{-4}^{3/10} ~ v_8^{1/10} ~ t_{10}^{2/5} ~ \mathrm{pc}.
    \label{eq:shock radius}
\end{equation}
The shock can act as a particle accelerator.
Particles propagate by advection and diffusion both upstream and downstream of the TS. We assume that the upstream advection velocity is constant in space $u_1(r)=u_1$, while the downstream velocity follows $u_2(r) = (u_1/4) (r/R_{\mathrm{s}})^{-2}$, where $u_1/4$ is the velocity immediately downstream. The diffusion coefficient in the bubble can be estimated by assuming a Kolmogorov or a Kraichnan phenomenology, in which cases:
\begin{equation}
    D_{\mathrm{Kol}}(E) \simeq \frac{1}{3} c L_{\mathrm{c}}^{2/3} r_{\mathrm{L}}^{1/3}, \ 
    \label{eq:kol}
\end{equation}
and
\begin{equation}
    D_{\mathrm{Kra}}(E) \simeq \frac{1}{3} c L_{\mathrm{c}}^{1/2} r_{\mathrm{L}}^{1/2}, 
    \label{eq:kra}
\end{equation}
respectively, with $r_{\mathrm{L}} = pc/(eB)$ being the Larmor radius. $L_{\mathrm{c}}$ is the coherence scale of the perturbations, which is assumed here to be of the same order of magnitude as the average separation between massive stars in the cluster, $L_{\mathrm{c}} \simeq \langle d \rangle$.
For Wd 1, which contains more than one hundred massive stars and has a half-mass radius of $R_\mathrm{sc} \sim 1$ pc \citep{Westerlund2022}, the average separation is estimated to be $\langle d \rangle \sim R_\mathrm{sc} N_\star^{-1/3} \lesssim 1$ pc.
Note that Equation \ref{eq:kol} and \ref{eq:kra} are valid only for energies such that $r_\mathrm{L}<L_\mathrm{c}$, while the scaling of the diffusion coefficient at higher energies has the following functional shape: $D(E) \simeq (1/3) c L_{\mathrm{c}} (r_{\mathrm{L}}/L_{\mathrm{c}})^{2}$ \cite[]{Subedi2017}.

Below we will also consider the rather extreme case of Bohm diffusion, $D_{\mathrm{Bohm}}\simeq (1/3) c r_{\mathrm{L}}$, keeping in mind however that it is physically meaningful only for those situations in which the turbulence is self-generated and the spectrum of the accelerated particles is close to the canonical $\propto E^{-2}$ (for relativistic particles). For star clusters it is not guaranteed that this may be the case (see discussion in \cite{Morlino_2021,Blasi_2023}).

The magnetic field in the bubble can be estimated by assuming that a fraction $\eta_B$ (<1) of the wind kinetic energy is converted into turbulent magnetic energy, i.e., $B^2/4\pi=\eta_B \rho u^2$, where $\rho$ is the mass density of the wind. Then the magnetic field immediately upstream of the termination shock can be estimated as:
\begin{equation}
    B_1(R_\mathrm{s}) \simeq 10.6 ~ \eta_B^{1/2} ~ \xi_{\mathrm{b}}^{1/5} ~ n_{10}^{3/10} ~ \dot{M}_{-4}^{1/5} ~ v_8^{2/5} ~ t_{10}^{-2/5} ~ \mu \mathrm{G}.
    \label{eq:magnetic field}
\end{equation}
For a strong shock, the magnetic field immediately downstream, $B_2(R_\mathrm{s})$, is enhanced by a factor of $\sqrt{11}$ \citep{Morlino_2021}. We assume that the Alfv\'enic Mach number $M_{\mathrm{A}}=v_{\mathrm{w}}/v_{\mathrm{A}}=1/\sqrt{\eta_B} \gtrsim$ 5, which sets a limit on $\eta_B \lesssim$ 0.04 if efficient particle acceleration at the termination shock can take place. Assuming that the turbulent magnetic field is generated locally, then it would have a spatial dependence on the radius, with $B_1(r) \propto r^{-1}$ in the upstream and $B_2(r) \propto r^{-2}$ in the downstream. The spatial dependence of the magnetic field is transferred to the diffusion coefficient through $r_{\mathrm{L}}$. 
Note that whether these diffusion coefficients are spatially constant or not depends on whether one assumes that turbulence is produced at some location and advected or rather generated at each location. In either case this is simply an assumption, since we are not following the cascading of turbulence in $k$-space here, nor its advection.

We numerically solve the transport equation for non-thermal particles with the same technique previously discussed by \cite{Blasi_2023}:
\begin{equation}
    \frac{1}{r^2} \frac{\partial}{\partial r} \left[r^2 D \frac{\partial f}{\partial r}\right] - \frac{1}{r^2} \frac{\partial}{\partial r} \left[r^2\tilde{u} f\right] + \frac{1}{p^2} \frac{\partial}{\partial p} \left[p^2\dot{p} f\right] + Q = \frac{\partial f}{\partial t},
\end{equation}
where $f(r,p,t)$ is the particle distribution function in phase space. The function $Q(r,p,t)$ represents the injection term, which can be expressed by $Q(r,p,t)=4 \pi R_{\mathrm{s}}^2 \eta_{\mathrm{inj}} n_1 u_1 [\delta(p - p_{\mathrm{inj}})/(4 \pi p^2)] [\delta(r - R_{\mathrm{s}})/(4 \pi r^2)]$. The energy loss term includes both the adiabatic energy loss/gain and the energy losses by interactions, namely $\dot{p}=\dot{p}_{\mathrm{ad}}+\dot{p}_{\mathrm{int}}$. For protons, we consider hadronic proton-proton (pp) interactions, while for electrons we include synchrotron radiation, inverse Compton scattering, and bremsstrahlung. $D(r,p)$ and $\tilde{u}(r)=u+\eta v_{\mathrm{A}}$ are the diffusion coefficient and the effective advection velocity, respectively. The coefficient $\eta$ mimics the effect of a net motion of the scattering centers, namely the possibility of non-vanishing local Alfv\'en speed \cite[]{Blasi_2023}.  Here we assume that waves propagate isotropically in the upstream ($\eta_1=0$), while anisotropy can exist in the downstream ($\eta_2\neq0$). The transport equation is solved numerically, with zero flux boundary condition at $r=0$ and free escape boundary at $r=R_\mathrm{b}$ (see Appendix \ref{app:transport} for details). We assume that particle transport has reached steady-state, which is justified for most of the cases in our calculations.

\section{Cluster environment}
\label{sec:env}

\begin{figure*}[t]
    \centering
    
    \begin{subfigure}{0.32\textwidth}
        \centering
        \includegraphics[width=\linewidth]{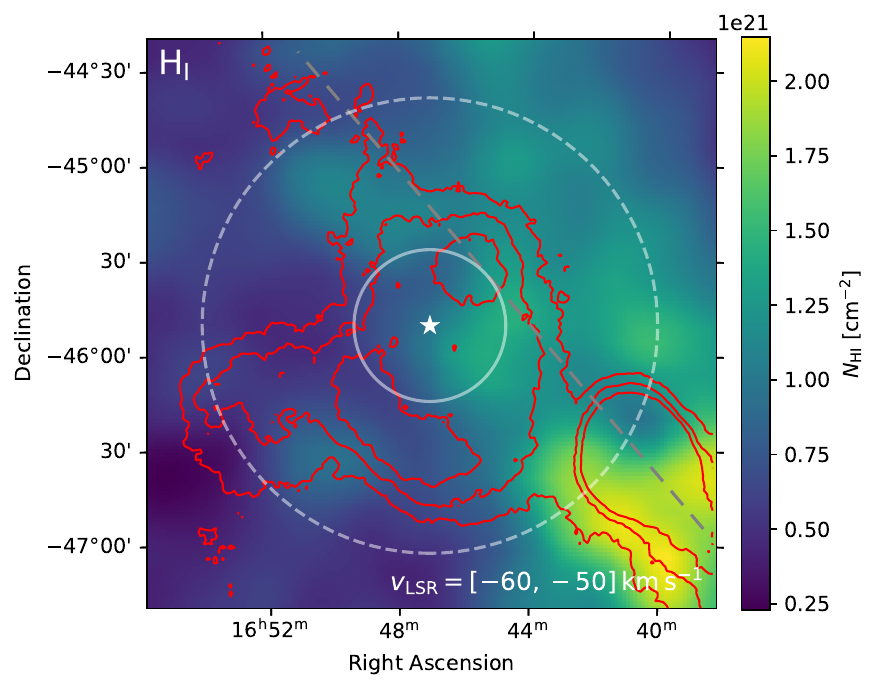}
    \end{subfigure}
    \begin{subfigure}{0.32\textwidth}
        \centering
        \includegraphics[width=\linewidth]{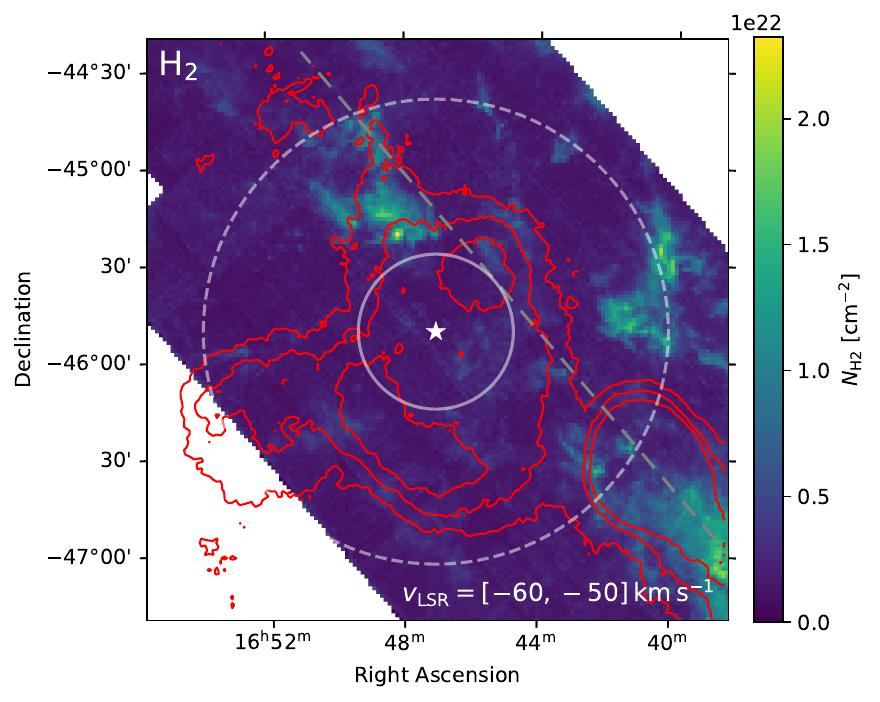}
    \end{subfigure}
    \begin{subfigure}{0.32\textwidth}
        \centering
        \includegraphics[width=\linewidth]{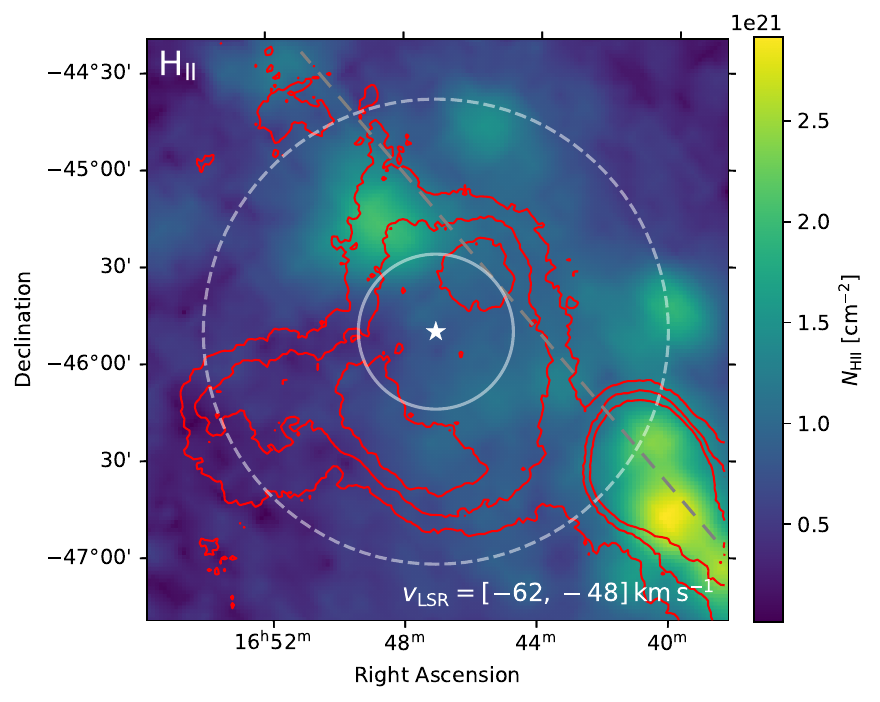}
    \end{subfigure}

    \begin{subfigure}{0.32\textwidth}
        \centering
        \includegraphics[width=\linewidth]{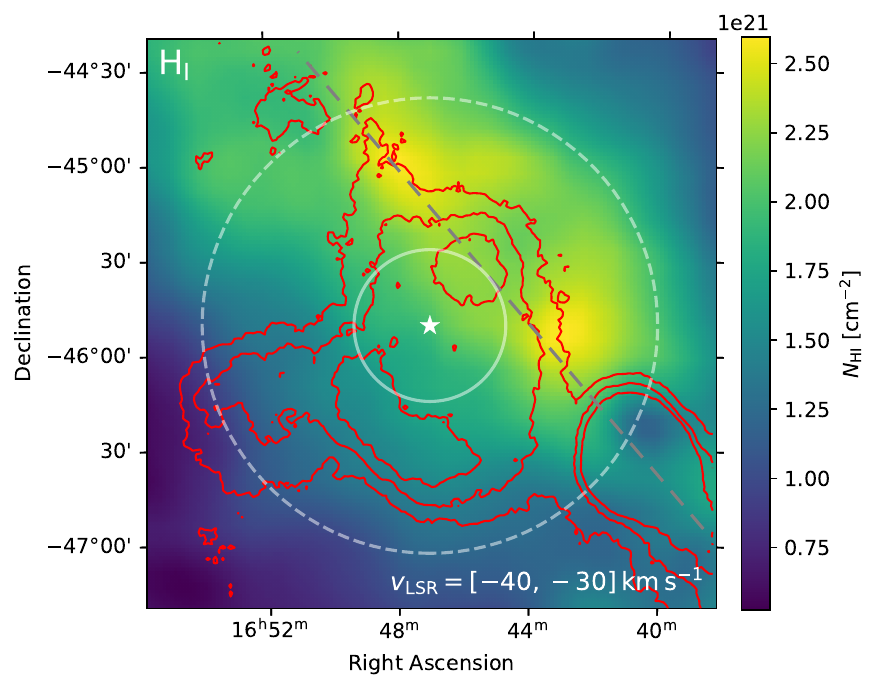}
    \end{subfigure}
    \begin{subfigure}{0.32\textwidth}
        \centering
        \includegraphics[width=\linewidth]{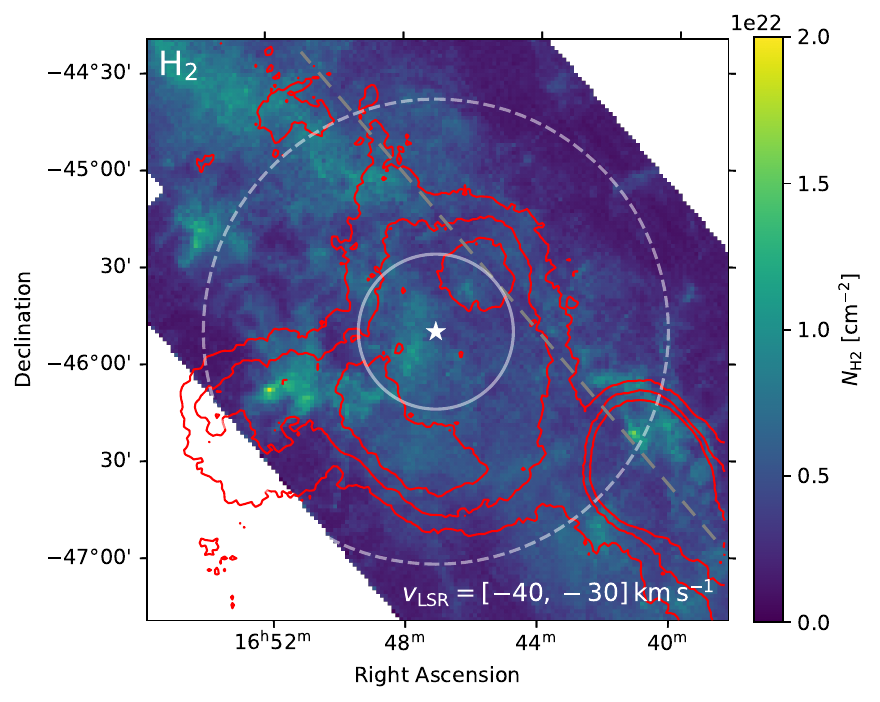}
    \end{subfigure}
    \begin{subfigure}{0.32\textwidth}
        \centering
        \includegraphics[width=\linewidth]{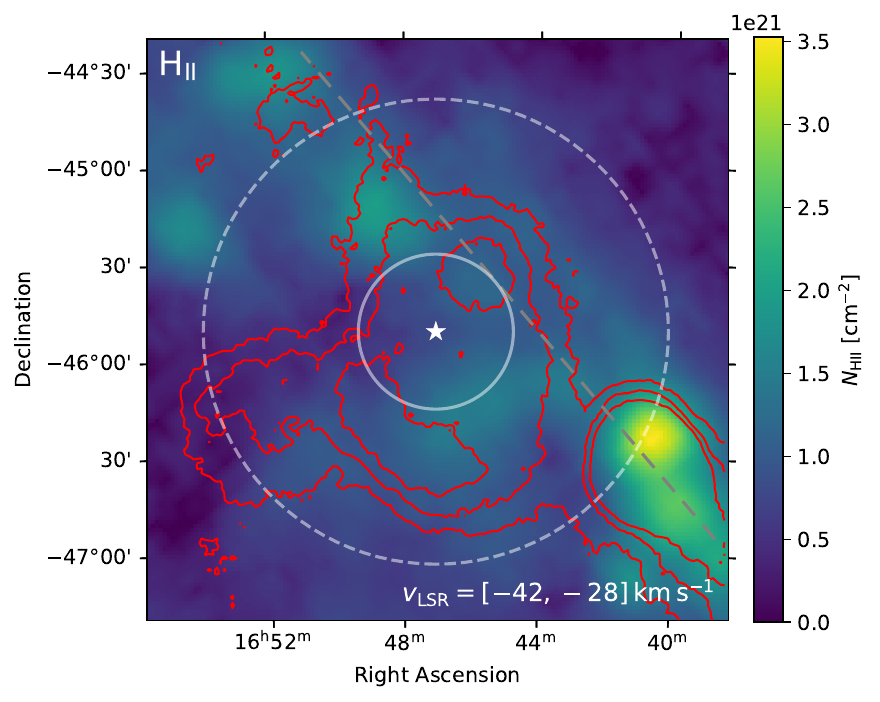}
    \end{subfigure}

    \caption{Column density maps for atomic hydrogen ($\mathrm{H}_\mathrm{I}$), molecular hydrogen ($\mathrm{H}_2$) and ionized hydrogen ($\mathrm{H}_\mathrm{II}$). A velocity cut $v_\mathrm{LSR}=[-60,-50] \, \mathrm{km} \, \mathrm{s}^{-1}$ is adopted for maps in the top, whereas $v_\mathrm{LSR}=[-40,-30] \, \mathrm{km} \, \mathrm{s}^{-1}$ in the bottom. For $\mathrm{H}_\mathrm{II}$, a slightly wider velocity range is adopted to ensure that it exceeds the velocity resolution of the data. The dashed circle ($\sim 1.2^\circ$) marks the region used to estimate the average gas density, while the solid circle indicates the location of the termination shock ($\sim 0.4^\circ$). The white star marks Wd 1, and the gray dashed line the Galactic plane. The red contours representing different gamma-ray flux levels are adopted from \cite{Westerlund2022}.}
    \label{fig:gas}
\end{figure*}

\subsection{Gas distributions}
\label{sec:env:gas}
We consider here three different components of interstellar gas in the vicinity of Wd 1: atomic hydrogen, molecular hydrogen and ionized hydrogen. The atomic hydrogen and molecular hydrogen can be traced by 21 cm line and CO ($J=1-0$) line, respectively. 

Before proceeding with this analysis, it is worth mentioning the work by \cite{Sano_2026ApJ..1004...12S}, where the authors find an association between gas clouds and HESS J1646--458. They reported the presence of two components of molecular clouds with different $v_\mathrm{LSR}$ at the same distance of Wd 1, and interpreted this as cloud-cloud collision. For this reason, we will perform the analysis for two different velocity ranges: $v_\mathrm{LSR}=[-60,-50] \, \mathrm{km} \, \mathrm{s}^{-1}$ and $[-40,-30] \, \mathrm{km} \, \mathrm{s}^{-1}$, corresponding to the same distance of $3.9 \pm 0.7 \, \mathrm{kpc}$ for Wd 1 \citep{Westerlund2022,Kothes_2007A&A...468..993K}. Furthermore, they suggested that the "HI-dip" associated with the peak of the southeastern gamma-ray shell is due to HI self-absorption \citep{Sano_2026ApJ..1004...12S}.

Here we use HI data from the Parkes Galactic All-Sky Survey third data release\footnote{\url{https://www.astro.uni-bonn.de/hisurvey/gass/}} (GASS III) \citep{GASS_2009,GASS_2015}. 
The intensity of HI emission is converted to column density $N_\mathrm{HI}$ using the conversion factor $X_\mathrm{HI}=1.823\times10^{18}\,\mathrm{cm}^{-2}/(\mathrm{K}\,\mathrm{km}\,\mathrm{s}^{-1})$ \citep{Aharonian_2019,Westerlund2022}. The column density map of HI is shown in Figure \ref{fig:gas}. For $v_\mathrm{LSR}=[-60,-50] \, \mathrm{km} \, \mathrm{s}^{-1}$, the total mass of HI in a circular region with radius of $\sim1.2^\circ$ centered on Wd 1 ($\sim 80 \, \mathrm{pc}$ at a distance of  $3.9 \, \mathrm{kpc}$) is estimated to be $M_\mathrm{HI}\approx1.59\times10^5M_\odot$, which translates into an average density of $n_\mathrm{HI}\approx2.8\,\mathrm{cm}^{-3}$ in the sphere, whereas for $v_\mathrm{LSR}=[-40,-30] \, \mathrm{km} \, \mathrm{s}^{-1}$, we obtain $M_\mathrm{HI}\approx3.11\times10^5M_\odot$ and $n_\mathrm{HI}\approx5.5\,\mathrm{cm}^{-3}$.

Similarly, to estimate the amount of molecular hydrogen in this region, we use data from the Mopra Southern Galactic Plane CO Survey\footnote{\url{https://doi.org/10.25919/9z4p-mj92}} \citep{Mopra_2018,Mopra_2023}, whose results are shown in Figure \ref{fig:gas}. The velocity cut is the same as that in the analysis of HI. The adopted CO-to-H$_2$ conversion factor is $X_\mathrm{CO}=1.5\times10^{20}\,\mathrm{cm}^{-2}/(\mathrm{K}\,\mathrm{km}\,\mathrm{s}^{-1})$ \citep{Westerlund2022}. The total mass of H$_2$ inside $1.2^\circ$ is $M_\mathrm{H2}\approx8.44\times10^5M_\odot$ and the number density is $n_\mathrm{H2}\approx7.5\,\mathrm{cm}^{-3}$ for $v_\mathrm{LSR}=[-60,-50] \, \mathrm{km} \, \mathrm{s}^{-1}$, whereas $M_\mathrm{H2}\approx1.48\times10^6M_\odot$ and $n_\mathrm{H2}\approx13.2\,\mathrm{cm}^{-3}$ for $v_\mathrm{LSR}=[-40,-30] \, \mathrm{km} \, \mathrm{s}^{-1}$.

In the downstream of the termination shock, the shocked wind forms a hot, rarefied bubble with temperature on the order of $\sim10^6$ K. In such environment, atomic hydrogen can be easily ionized. The number density of the shocked wind in the bubble can be estimated as (see Equation \ref{eq:n2}):
\begin{equation}
    n_\mathrm{w} = \frac{3\dot M T_\mathrm{age}}{4 \pi R_{\mathrm{b}}^3 m_\mathrm{H}},
    \label{eq:nw}
\end{equation}
which is $\sim 0.02 \, \mathrm{cm}^{-3}$ for typical parameters of Wd 1. On the other hand, cold, dense material accumulated in the outer shell of the hot bubble, as well as clumpy molecular clouds within the bubble, can be evaporated through thermal conduction, thereby contributing to the ionized gas content of the bubble. The mass evaporation rate of the shell can be expressed as (see Equation \ref{eq:mass gain rate}) \citep{Castor_1975,Menchiari_2023}:
\begin{equation}
    \dot M_\mathrm{b} = \frac{16\pi\mu}{25k_\mathrm{B}}  K(\tilde{T}_2) R_\mathrm{b},
    \label{eq:Mev}
\end{equation}
where $K(\tilde{T}_2) = C \tilde{T}_2^{5/2}$ is the thermal conductivity, with $C=1.2\times10^{-6}\,\mathrm{erg}\,\mathrm{cm}^{-1}\,\mathrm{s}^{-1}\,\mathrm{K}^{-7/2}$. The gas temperature immediately downstream of the termination shock, $\tilde{T}_2$, including the effect of thermal conduction, can be written as (see Equation \ref{eq:tildeT2}):
\begin{equation}
    \tilde{T}_2\simeq2.5\times10^6 ~ \xi_{\mathrm{b}}^{8/35} ~ n_{10}^{2/35} ~ \dot{M}_{-4}^{8/35} ~ v_8^{16/35} ~ t_{10}^{-6/35} ~ \mathrm{K}.
\end{equation}
Here we adopt the mean molecular weight $\mu\approx0.62 m_\mathrm{H}$. Substituting Equation \ref{eq:Mev} into \ref{eq:nw} we find the number density of ionized gas evaporated from the shell, $n_\mathrm{ev}\sim 0.2 \, \mathrm{cm}^{-3}$ for typical parameters of Wd 1. Therefore, ionized gas is mainly produced by evaporation of the shell, and the total number density of the ionized gas is $n_\mathrm{HII}=n_\mathrm{w}+n_\mathrm{ev} \approx n_\mathrm{ev}$ \citep{Peron_2025}.

To compare the estimated ionized gas density with observations, we use radio recombination line (RRL) data obtained from the HI Parkes All-Sky Survey\footnote{\url{https://www.jb.man.ac.uk/research/parkes_rrl_survey/}} (HIPASS) \citep{HIPASS_1996,HIPASS_2015} and associated Zone of Avoidance (ZOA) Survey \citep{ZOA_1998}. We convert the velocity-integrated intensity of RRL emission into the emission measure ($\mathrm{EM}=\int\,n_\mathrm{e}^2\,dl$, in units of $\mathrm{pc}\,\mathrm{cm}^{-6}$) through a conversion factor of $1.74\times10^{-3}\,T_\mathrm{e}^{1.5}\,\nu_\mathrm{GHz}$, where $T_\mathrm{e}\approx6000\,\mathrm{K}$ is the electron temperature at the location of Wd 1 and $\nu_\mathrm{GHz}=1.4$ in units of GHz is the rest frequency of the RRL emission \citep{HIPASS_2015}. The total EM calculated from RRL emission is consistent with that obtained from the Planck free-free map \citep{Planck_2016}. Assuming that the ionized gas is uniformly distributed in a slab perpendicular to the line of sight, with a thickness of $L\sim 160\,\mathrm{pc}$ and centered on Wd 1, the column density can be estimated as $N_\mathrm{HII}=\sqrt{EM\times L}$. The results are shown in Figure \ref{fig:gas}. For $v_\mathrm{LSR}=[-62,-48] \, \mathrm{km} \, \mathrm{s}^{-1}$, the total mass of ionized gas in the dashed circle is $M_\mathrm{HII}\approx1.5\times10^5M_\odot$, corresponding to $n_\mathrm{HII}\approx1.8\,\mathrm{cm}^{-3}$ in the cylinder, whereas for $v_\mathrm{LSR}=[-42,-28] \, \mathrm{km} \, \mathrm{s}^{-1}$, we obtain $M_\mathrm{HII}\approx1.6\times10^5M_\odot$, and $n_\mathrm{HII}\approx1.9\,\mathrm{cm}^{-3}$. The observed $n_\mathrm{HII}$ is about one order of magnitude higher than the analytical value estimated above, indicating the evaporation of additional material in this region (e.g., clumpy molecular cloud in the bubble \citep{Blasi_2023}) .

The analysis above shows that the total average number density of the gas around Wd 1 is on the order of $10\,\mathrm{cm}^{-3}$. For atomic and molecular hydrogen gas, the number density in the velocity range $v_\mathrm{LSR}=[-40,-30] \, \mathrm{km} \, \mathrm{s}^{-1}$ is about twice of that in $[-60,-50] \, \mathrm{km} \, \mathrm{s}^{-1}$, whereas for ionized hydrogen, the number densities are similar for the two velocity ranges.

\subsection{Magnetic field and photon fields}

As described in Section \ref{sec:model}, we assume that the magnetic energy density is a fraction of the local plasma kinetic energy, $B^2/4\pi=\eta_B \rho u^2$. In the upstream region, $\rho_1\propto r^{-2}$ and $u_1$ is constant; therefore, the magnetic field scales as $B_1(r)\propto r^{-1}$. In the downstream, $\rho_2$ is constant and $u_2\propto r^{-2}$, yielding $B_2(r)\propto r^{-2}$. Thus, the magnetic field in the bubble can be expressed as:
\begin{equation}
    B(r) = 
    \begin{cases}
    B_1(R_\mathrm{s}) \left(\frac{r}{R_{\mathrm{s}}}\right)^{-1}, & R_{\mathrm{sc}}<r \leq R_{\mathrm{s}} \\
    B_2(R_\mathrm{s}) \left(\frac{r}{R_{\mathrm{s}}}\right)^{-2}, & R_{\mathrm{s}}<r \leq R_{\mathrm{b}}
    \end{cases}
\end{equation}
where $B_1(R_\mathrm{s})$ and $B_2(R_\mathrm{s})$ are the magnetic fields immediately upstream and downstream, with $B_2(R_\mathrm{s})=\sqrt{11}B_1(R_\mathrm{s})$ (see Equation \ref{eq:magnetic field}). $R_{\mathrm{sc}} \simeq 1\, \mathrm{pc}$ is the size of the star cluster. 

The photon fields considered in this work include the cosmic microwave background (CMB, $T_{\mathrm{CMB}}=2.73 \, \mathrm{K}$, $u_{\mathrm{CMB}}=0.26 \, \mathrm{eV} \, \mathrm{cm}^{-3}$), the far-infrared radiation (FIR, $T_{\mathrm{FIR}}=35 \, \mathrm{K}$, $u_{\mathrm{FIR}}=0.6 \, \mathrm{eV} \, \mathrm{cm}^{-3}$) and the near-infrared radiation (NIR, $T_{\mathrm{NIR}}=3000 \, \mathrm{K}$, $u_{\mathrm{NIR}}=1.2 \, \mathrm{eV} \, \mathrm{cm}^{-3}$) \citep{Popescu_2017}. The ultraviolet (UV) component is not important and is neglected here. 

In addition to these photon fields, we also account for the radiation field from the collection of massive stars in Wd 1. The energy density of the cluster photon field at the termination shock can be estimated as $u_{\mathrm{SC}}(R_\mathrm{s})=L_{\mathrm{bol}}/(4 \pi R_\mathrm{s}^2c) \approx 20.6 (L_{\mathrm{bol}}/10^{41} \, \mathrm{erg} \, \mathrm{s}^{-1}) (R_\mathrm{s}/29.1 \, \mathrm{pc})^{-2} \, \mathrm{eV} \, \mathrm{cm}^{-3}$, where $L_{\mathrm{bol}} \approx 10^{41} \, \mathrm{erg} \, \mathrm{s}^{-1}$ is the bolometric luminosity of Wd 1 \citep{Harer_2023}. Thus, the energy density of the cluster photon filed at each location can be expressed as:
\begin{equation}
    u_\mathrm{SC}(r) = 
    u_{\mathrm{SC}}(R_\mathrm{s}) \left(\frac{r}{R_{\mathrm{s}}}\right)^{-2}, \quad R_{\mathrm{sc}} \leq r \leq R_{\mathrm{b}}\ .
\end{equation}
We assume an effective temperature of the cluster, $T_\mathrm{eff} \sim 30000\, \mathrm{K}$, typical of OB-type stars \citep{Martins_2005A&A...436.1049M, Crowther_2006A&A...446..279C}. We note that Wd 1 hosts at least 24 Wolf-Rayet stars and more than 100 OB giants and supergiants \citep{Clark_2020, Crowther_2006MNRAS.372.1407C}, whose effective temperature may span a wide range. However, variations in the effective temperature have little impact on our results, since the IC component of the cluster photon field is subdominant compared to those of the CMB and FIR fields, as we will show below.

\section{Leptonic model}
\label{sec:lep}

\begin{figure*}[t]
    \centering
    
    \begin{subfigure}{0.9\textwidth}
        \centering
        \includegraphics[width=\linewidth]{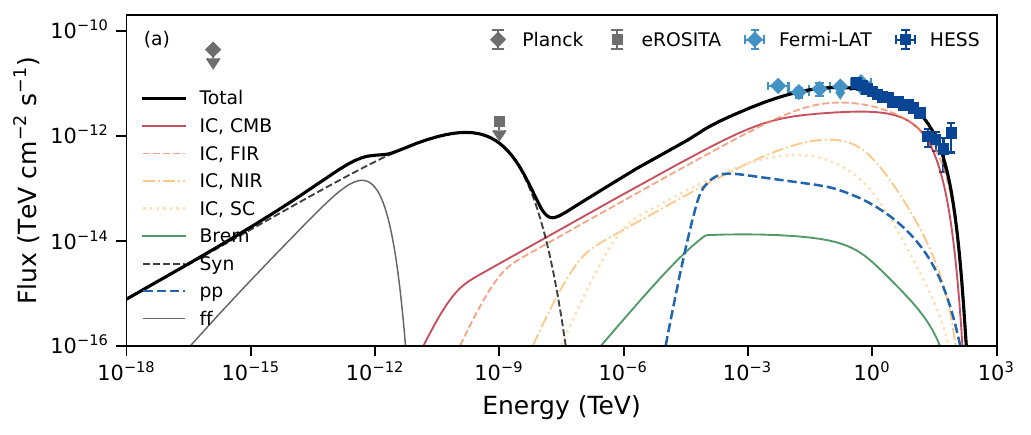}
    \end{subfigure}
    
    \begin{subfigure}{0.45\textwidth}
        \centering
        \includegraphics[width=\linewidth]{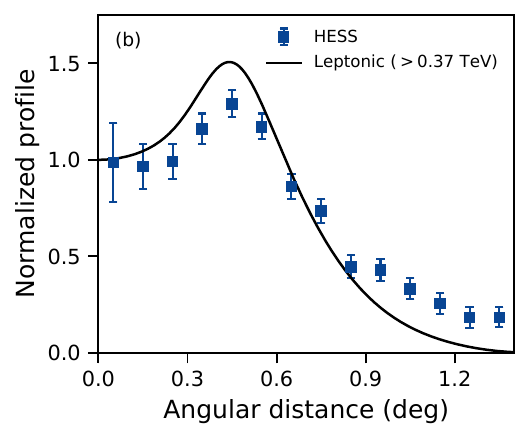}
    \end{subfigure}
    \begin{subfigure}{0.45\textwidth}
        \centering
        \includegraphics[width=\linewidth]{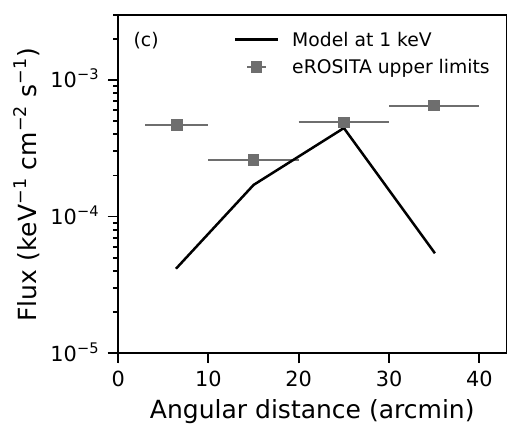}
    \end{subfigure}

    \caption{Leptonic model with Bohm diffusion. (a) Spectral energy distribution. The warm lines represent the IC fluxes for different photon fields. The green solid, gray solid and black dashed lines are bremsstrahlung, free-free and synchrotron fluxes, respectively. The blue dashed line is the hadronic gamma-ray flux. The light blue diamonds and dark blue squares are Fermi-LAT and HESS data, whereas the gray diamond and gray square representing the Planck and eROSITA upper limits. (b) Gamma-ray intensity morphology. The solid line is the leptonic gamma-ray intensity, normalized to the value at the center. (c) X-ray flux morphology. The solid line is the predicted X-ray flux morphology at 1 keV.}
    \label{fig:lep_bohm}
\end{figure*}

\begin{figure*}[t]
    \centering
    
    \begin{subfigure}{0.9\textwidth}
        \centering
        \includegraphics[width=\linewidth]{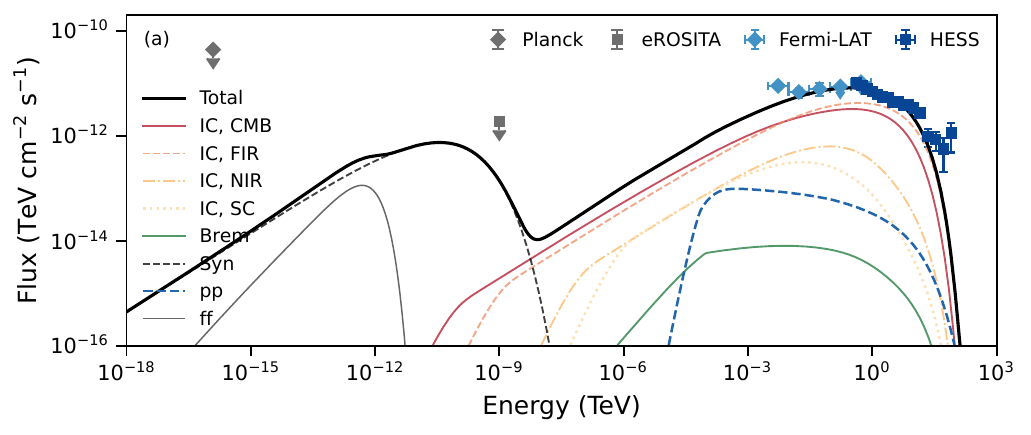}
    \end{subfigure}
    
    \begin{subfigure}{0.45\textwidth}
        \centering
        \includegraphics[width=\linewidth]{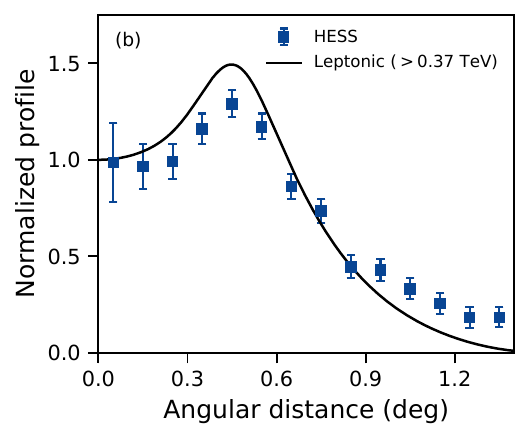}
    \end{subfigure}
    \begin{subfigure}{0.45\textwidth}
        \centering
        \includegraphics[width=\linewidth]{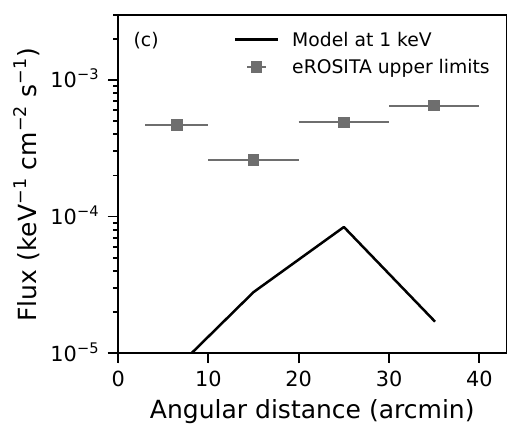}
    \end{subfigure}

    \caption{Same as Figure \ref{fig:lep_bohm}, but for Kraichnan diffusion.}
    \label{fig:lep_kra}
\end{figure*}

Previous studies favor the leptonic scenario for interpreting the gamma-ray spectrum from Wd 1, owing to the lack of association between the gamma-ray emission and molecular clouds or atomic gas in this region \citep{Westerlund2022,Harer_2023,Lemoine-Goumard_2025NatCo..1610820L}. However, these works did not engage in a detailed analysis of the gamma-ray morphology measured by HESS, which is crucial to gather valuable information on electron transport and energy losses in the bubble. Here, we revisit the leptonic scenario by using the model described in Section \ref{sec:model}, and simultaneously study the gamma-ray spectrum and morphology of Wd 1. We calculate the gamma-ray emission from inverse Compton (IC) scattering and bremsstrahlung radiation of relativistic electrons in the environment described in Section \ref{sec:env}. Synchrotron emission is also calculated and compared with the X-ray upper limit of eROSITA \citep{erosita_2025}. 
In standard DSA theory, both electrons and protons can be accelerated at the shock. Therefore, we also include the corresponding hadronic component from pp interactions of these protons with the ambient gas, assuming a standard electron-to-proton ratio $K_\mathrm{ep} \sim 0.01$ \citep{Strong_2007,Park_2015PhRvL.114h5003P}. 
Here we provisionally assume that the gas density in the bubble is that of the shocked wind, $n_\mathrm{b} = n_\mathrm{w} \sim 0.02 \, \mathrm{cm}^{-3}$ (see Equation \ref{eq:nw}), although the average density inferred from observations in the vicinity of Wd 1 is of order $\sim 10 \, \mathrm{cm}^{-3}$, as discussed in Sec.\ref{sec:env:gas}. Therefore our estimate of the hadronic contribution to the gamma-ray flux from Wd 1 should be regarded as very conservative.

Figure \ref{fig:lep_bohm}(a) shows the predicted spectral energy distribution (SED) for the leptonic model with Bohm diffusion. The model parameters are listed in Table \ref{tab:para}. It is obvious that the gamma-ray flux is dominated by the IC emission (lines with warm colors) of relativistic electrons, in which the IC component from CMB is comparable to that from FIR, while the IC components from NIR and SC are apparently lower. The corresponding hadronic gamma-ray flux (blue dashed line) is negligible due to the low density of the shocked wind. 
The GeV data measured by Fermi-LAT (light blue diamonds \citep{Lemoine-Goumard_2025NatCo..1610820L}) and the TeV data measured by HESS (dark blue squares \citep{Westerlund2022, Lemoine-Goumard_2025NatCo..1610820L}) are well fitted, except for the first and last data points. 
The adopted acceleration efficiency for electron is $\xi_\mathrm{CR,\,e} \approx 10^{-3}$, and the corresponding acceleration efficiency for proton is $\xi_\mathrm{CR,\,p} \approx 0.1$ for a given electron-to-proton ratio $K_\mathrm{ep} = 0.01$ at the shock (for particle spectrum with a slope of -2). It is unlikely to increase the hadronic component using a larger $\xi_\mathrm{CR,\,p}$ because the acceleration efficiency is thought to be $\lesssim 0.1$ considering the energetics of CRs in the Milky Way. On the other hand, if, in order to be conservative, we only include the gas density from the evaporated gas, i.e., $n_\mathrm{b} = n_\mathrm{w} + n_\mathrm{ev} \sim 0.2 \, \mathrm{cm}^{-3}$, the hadronic contribution increases just by a factor of $\sim 10$ compared to the case of wind alone. A larger value of the density, while allowed by observations, would make the hadronic contribution dominant, as discussed in Section \ref{sec:had}.

The synchrotron emission within $40^\prime$ of Wd 1 is also shown and compared with the radio upper limit from Planck satellite \citep{Planck_2020A&A...641A...2P} and the X-ray upper limit from the eROSITA telescope \citep{erosita_2025}, as seen in Figure \ref{fig:lep_bohm}(a). The average radio intensity within $1^\circ$ around Wd 1 is derived by \cite{Westerlund2022}, and rescaled to $40^\prime$ here in order to compare with our model ($\Phi_{\mathrm{30\,\mathrm{GHz}, \, u.l.}} \sim 7 \times 10^{-11} \, \mathrm{erg} \, \mathrm{cm}^{-2} \, \mathrm{s}^{-1}$). \cite{erosita_2025} analyzed the spectra of four annuli around Wd 1 that coincide with HESS J1646--458 to search for synchrotron emission. They found no evidence of synchrotron emission and derived an upper limit of the X-ray flux up to $40^\prime$ around Wd 1, $\Phi_{\mathrm{1\,\mathrm{keV}, \, u.l.}} \sim 3 \times 10^{-12} \, \mathrm{erg} \, \mathrm{cm}^{-2} \, \mathrm{s}^{-1}$ for $\Gamma=2.5$. When the gamma-ray flux is fitted with IC emission, the X-ray upper limit constrains the magnetic field immediately downstream of the termination shock to $\lesssim 4 \, \mu \mathrm{G}$, which is consistent with the estimations from previous works \citep{Harer_2023}.
In addition, we also compute the free-free emission (thermal bremsstrahlung) from the thermal bubble following \cite{Draine_2011piim.book.....D}. We assume that the gas is fully ionized and adopt an electron temperature of 6000 K\footnote{Here we adopt the electron temperature corresponding to the Galactocentric radius of Wd 1, rather than the temperature expected for the hot bubble.}.

In Figure \ref{fig:lep_bohm}(b) we show the gamma-ray morphology based on our calculations, compared with the HESS data points corresponding to events with energy above 0.37 TeV, normalized to the intensity at the center of the source. 
The solid line in Figure \ref{fig:lep_bohm}(b) shows the predicted gamma-ray morphology from electrons. The hadronic component is not plotted here, due to its negligible contribution to the total gamma-ray flux. 
The predicted morphology is convolved with the point spread function (PSF) of HESS with $\sigma_\mathrm{psf} \sim 0.08^\circ$ \citep{HESS_2018A&A...612A...1H}. It shows that our model reproduces the overall shape of the observed data, although some discrepancies remain near the peak and at large radii. We note that the gamma-ray morphology measured by HESS exhibits substantial variations across the different energy ranges and spatial segments defined in \cite{Westerlund2022}. Therefore, the discrepancies between our model and the data may be alleviated when considering these large variations.

Figure \ref{fig:lep_bohm}(c) shows the X-ray morphology at 1 keV. The eROSITA upper limits are derived from the four annuli up to $40^\prime$ as defined in \cite{erosita_2025}. Our predicted X-ray morphology is consistent with the upper limits in each annulus.

We also do the same calculations for the case of Kraichnan diffusion, the results of which are shown in Figure \ref{fig:lep_kra}. The maximum energy of electrons can be estimated by equating the acceleration and energy loss timescales. In the present study, this condition is roughly satisfied for $\eta_B\sim10^{-3}$. In order to obtain a larger $E_\mathrm{max}$, we adopt a smaller $L_\mathrm{c} = 0.2~\mathrm{pc}$. As can be seen in Figure \ref{fig:lep_kra}(a), although this model reproduce the gamma-ray flux below $\sim 10$ TeV, it fails to account for the highest-energy data points, due to severe energy losses of electrons. It is worth mentioning that the Galactic diffuse emission is not included in our calculation, but its contribution should be small considering the low gas density adopted here. Therefore, explaining the gamma-ray data with leptonic model requires Bohm diffusion, which is, however, unlikely to be fulfilled in this environment \citep{Morlino_2021,Blasi_2023}.

\begin{table*}
\caption{Model parameters.}
\label{tab:para}
\centering
\begin{tabular}{lccccccccccc}
\hline\hline
 & $L_\mathrm{c}$ & $R_\mathrm{s}$ & $R_\mathrm{b}$ & $n_0$ & $n_\mathrm{b}$ & $\eta_B$ & $B_2$ & $\eta_2$ & $\xi_\mathrm{CR, \, e}$ & $\xi_\mathrm{CR, \, p}$ & $K_\mathrm{ep}$ \\
 & ($\mathrm{pc}$) & ($\mathrm{pc}$) & ($\mathrm{pc}$) & ($\mathrm{cm}^{-3}$) & ($\mathrm{cm}^{-3}$) & & ($\mu \mathrm{G}$) & & & \\
\hline
\multicolumn{11}{c}{Spatially dependent Bohm} \\
\hline
Leptonic model & 1.0 & 29.1 & 94.7 & 30 & 0.02 & $1.0\times10^{-3}$ & 3.3 & 0.4 & $1.0\times10^{-3}$ & 0.1 & 0.01 \\
Hadronic model & 1.0 & 25.0 & 85.5 & 50 & 2 & 0.01 & 12.1 & 0 & $6.2\times10^{-5}$ & 0.062 & $10^{-3}$ \\
\hline
\multicolumn{11}{c}{Spatially dependent Kraichnan} \\
\hline
Leptonic model & 0.2 & 30.8 & 98.2 & 25 & 0.02 & $1.0\times10^{-3}$ & 3.1 & 0 & $7.0\times10^{-4}$ & 0.07 & 0.01 \\
Hadronic model & 0.2 & 25.0 & 85.5 & 50 & 2 & 0.04 & 24.2 & 0 & $8\times10^{-5}$ & 0.08 & $10^{-3}$ \\
\hline
\end{tabular}
\end{table*}

\section{Hadronic model}
\label{sec:had}

\begin{figure*}[t]
    \centering
    
    \begin{subfigure}{0.9\textwidth}
        \centering
        \includegraphics[width=\linewidth]{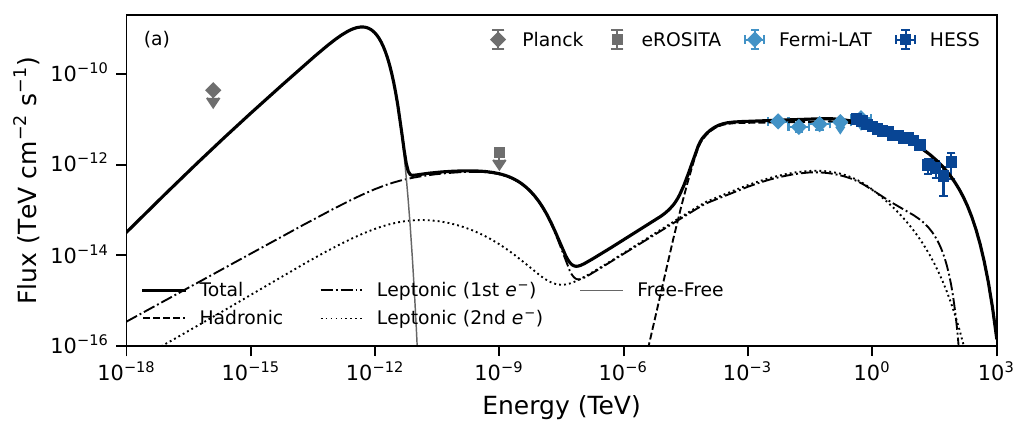}
    \end{subfigure}
    
    \begin{subfigure}{0.45\textwidth}
        \centering
        \includegraphics[width=\linewidth]{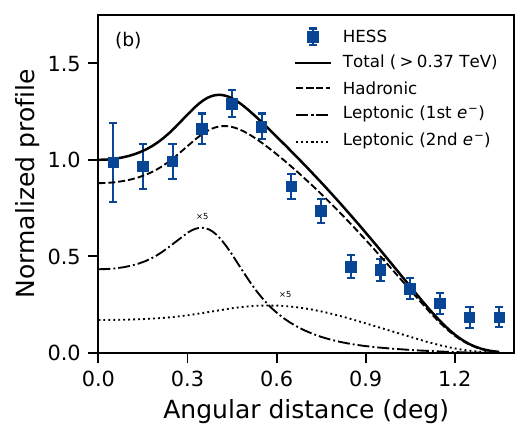}
    \end{subfigure}
    \begin{subfigure}{0.45\textwidth}
        \centering
        \includegraphics[width=\linewidth]{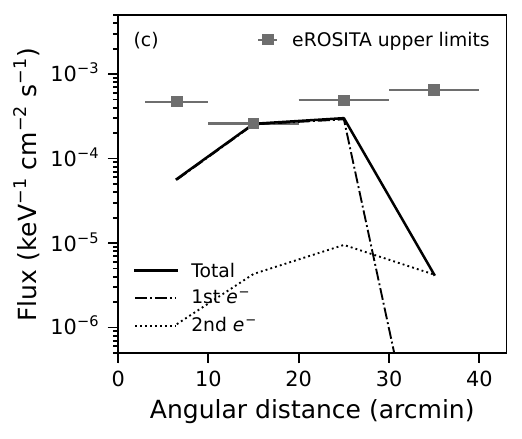}
    \end{subfigure}

    \caption{Hadronic model with Bohm diffusion. (a) Spectral energy distribution. The black dashed line represents the hadronic gamma-ray flux. The dashed-dotted line and dotted line are the total fluxes from primary electrons and secondary electrons. (b) Gamma-ray intensity morphology. The solid line is the sum of the leptonic and hadronic gamma-ray intensities, normalized to the value at the center. We multiply the gamma-ray intensities of primary and secondary electrons by a numerical factor for better clarifications. (c) X-ray flux morphology. The solid line is the sum of the predicted flux from primary and secondary electrons.}
    \label{fig:had_bohm}
\end{figure*}

\begin{figure*}[t]
    \centering
    
    \begin{subfigure}{0.9\textwidth}
        \centering
        \includegraphics[width=\linewidth]{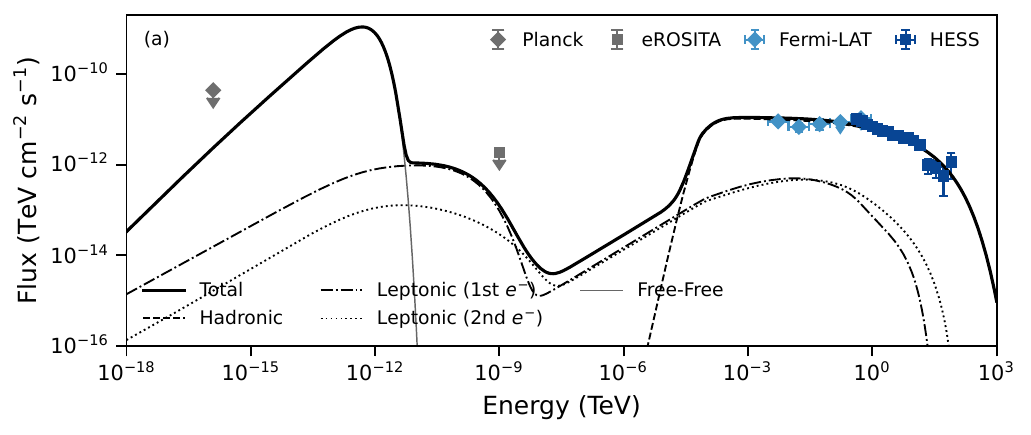}
    \end{subfigure}
    
    \begin{subfigure}{0.45\textwidth}
        \centering
        \includegraphics[width=\linewidth]{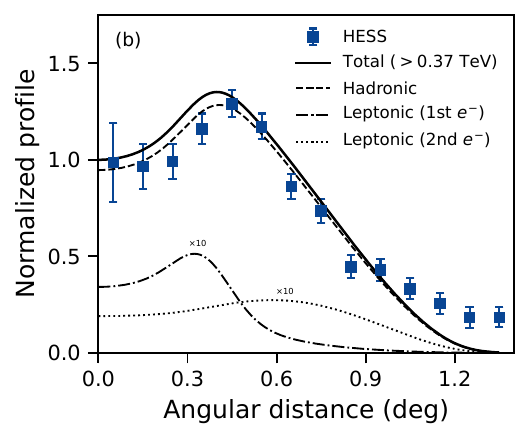}
    \end{subfigure}
    \begin{subfigure}{0.45\textwidth}
        \centering
        \includegraphics[width=\linewidth]{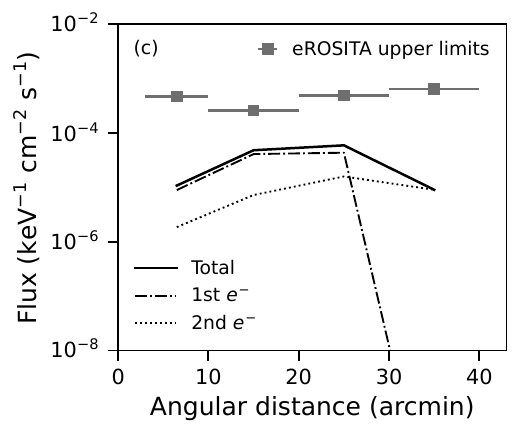}
    \end{subfigure}

    \caption{Same as Figure \ref{fig:had_bohm}, but for Kraichnan diffusion.}
    \label{fig:had_kra}
\end{figure*}

In the previous section, we have shown that the hadronic emission contributes negligibly ($\sim 100$ times lower) to the gamma-ray flux if the gas density in the bubble is the one of the shocked wind alone, $n_\mathrm{b} = n_\mathrm{w} \sim 0.02 \, \mathrm{cm}^{-3}$, and the proton acceleration efficiency is $\xi_\mathrm{CR, \, p}=0.1$.  Here, we investigate an hadronic scenario, where the gamma-ray flux is predominantly due to pp collisions. In order to make hadronic emission the dominant component while keeping $\xi_\mathrm{CR, \, p} \lesssim 0.1$, the gas density in the bubble should be $n_\mathrm{b} \gtrsim 2 \, \mathrm{cm}^{-3}$, which is what will be used in the following calculations. A similar value ($\sim 1.1 \, \mathrm{cm}^{-3}$) is adopted in \cite{Shi_2026JHEAp..5100560S}. We stress again that the total gas density inferred from observations is $\sim 10 \, \mathrm{cm}^{-3}$; therefore, $n_\mathrm{b} \gtrsim 2 \, \mathrm{cm}^{-3}$ is a reasonable value.
We use the same algorithm described before to solve the particle transport equations for protons and primary electrons. In addition to the primary electrons accelerated at the termination shock, we also account for the secondary electrons that are produced from pp interactions. For simplicity, we neglect the possible reacceleration of secondary electrons that reach the termination shock and only consider their transport and interactions in the downstream region. In this case, we adopt the zero flux requirement at $R_\mathrm{s}$ as the inner boundary condition. 
The results are shown in Figure \ref{fig:had_bohm} for Bohm diffusion and Figure \ref{fig:had_kra} for Kraichnan diffusion, with the model parameters listed in Table \ref{tab:para}.

Figure \ref{fig:had_bohm}(a) shows the SED of Wd 1. The black dashed, dashed-dotted and dotted lines represent the emission from protons, primary electrons and secondary electrons, respectively, in the case of Bohm diffusion in the bubble. The thick solid line is the sum of these components. Here we choose a smaller $K_\mathrm{ep}=10^{-3}$ to ensure both that the hadronic emission is the dominant one and that the X-ray emission does not violate the eROSITA upper limit. 
It is clear that once $n_\mathrm{b}$ is fixed, reproducing the gamma-ray flux as the result of hadronic interactions fixes $\xi_\mathrm{CR, \, p}$ and constrains $K_\mathrm{ep}$.
For $n_\mathrm{b} = 2 \, \mathrm{cm}^{-3}$, the adopted proton acceleration efficiency is $\xi_\mathrm{CR, \, p} \approx 0.06$; therefore, $n_\mathrm{b} \gtrsim 1.2 \, \mathrm{cm}^{-3}$ is required to keep $\xi_\mathrm{CR, \, p} \lesssim 0.1$ if the hadronic scenario operates alone. 

Notice that, while in terms of gamma-ray flux (at $\sim 1 \mathrm{TeV}$), primary and secondary electrons contribute comparably, in terms of synchrotron emission the former are largely dominant. This is due to the fact that secondary electrons are produced everywhere in the bubble, while primary electrons are injected and accelerated at the shock, near which the magnetic field is appreciably higher. 
We also note that the free-free flux is much higher than that in the leptonic case, since free-free emissivity is proportional to $n_e^2$. Here we assume that the gas within the bubble is fully ionized, which is reasonable since the ionized gas density obtained from our data analysis is of order $\sim 1.8 \, \mathrm{cm}^{-3}$ (see Section \ref{sec:env:gas}).

Figure \ref{fig:had_bohm}(b) shows the gamma-ray morphology. The HESS data points can be well reproduced by the hadronic model. As is also evident that the gamma-ray morphology of secondary electrons is much broader than that of primary electrons, due to the same reason explained above. Figure \ref{fig:had_bohm}(c) shows the predicted X-ray flux profile, which is mainly produced by primary electrons within an angular radius of $\sim 30^\prime$ from the center. 

Similarly, the results for Kraichnan diffusion are presented in Figure \ref{fig:had_kra}, and the model parameters are shown in Table \ref{tab:para}. We adopt $\eta_B=0.04$ and $L_\mathrm{c}=0.2~\mathrm{pc}$ here. As shown in Figure \ref{fig:had_kra}, the observational data can be well reproduced in this scenario.

\subsection{Neutrino detectability}

\begin{figure}[t]
    \centering
    \includegraphics[width=\linewidth]{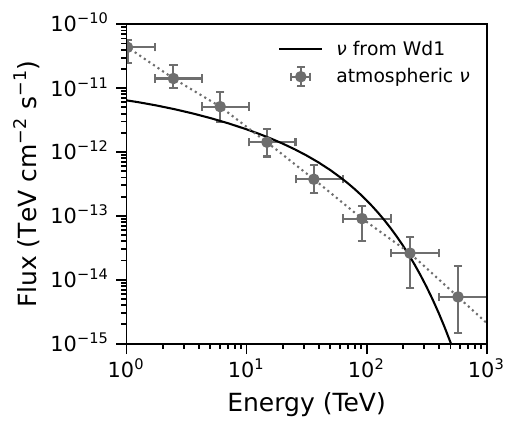}

    \caption{Predicted neutrino flux corresponding to Figure \ref{fig:had_bohm}.}
    \label{fig:nu}
\end{figure}

Nuclear collisions also produce neutrinos, which may serve as a smoking-gun signature of hadronic processes. Here we calculate the corresponding neutrino flux in the case of Bohm diffusion, as shown in Figure \ref{fig:nu}. The solid line represents the predicted flux of the all-flavor neutrinos from Wd 1.
We adopt the atmospheric neutrino intensity $I_{\nu_{\mu}, \, \mathrm{atm}}$ measured by IceCube \citep{IceCube_2015EPJC...75..116A,Mascaretti_2020APh...114...22M} to derive the atmospheric neutrino flux within the angular extent of Wd 1. The gray data points in Figure~\ref{fig:nu} show the derived flux, while the dotted line represents an interpolation.
The observed angular extent is defined as $\theta_{\mathrm{obs}} = (\theta_{\mathrm{s}}^2 + \theta_{\mathrm{psf}}^2)^{1/2}$, where $\theta_{\mathrm{s}}$ is the angular size of the gamma-ray emission of Wd 1 ($\sim 1.2^\circ$) and $\theta_{\mathrm{psf}}$ the angular resolution of KM3NeT ($\sim 0.2^\circ$ for track events above 10 TeV \citep{km3net_2016JPhG...43h4001A,km3net_2024EPJC...84..885K}). We choose KM3NeT because it is sensitive to sources in the southern sky where Wd 1 is located ($\delta \sim -46.26^\circ$).  
Figure~\ref{fig:nu} shows that the astrophysical neutrino flux is higher than the background flux in the energy range between 20 and 100 TeV.

We estimate the statistical significance of neutrinos from Wd 1 with KM3NeT for a given observation time:
\begin{equation}
    \sigma = \frac{N_\mathrm{s}}{\sqrt{N_\mathrm{b}}},
\end{equation}
where $N_\mathrm{s}$ and $N_\mathrm{b}$ are the numbers of source events and background events, respectively, which can be estimated as \citep{Gagliardini_2024ApJ...969..161G}:
\begin{equation}
    N_\mathrm{s} = T_\mathrm{obs} \int_{1 \, \mathrm{TeV}}^{{500 \, \mathrm{TeV}}} \, \frac{dN_\nu}{dE_\nu} \, A_\mathrm{eff}(E_\nu) \, V(\delta) \, dE_\nu,
\end{equation}
and
\begin{equation}
    N_\mathrm{b} = \Delta \Omega \, T_\mathrm{obs} \int_{1 \, \mathrm{TeV}}^{{500 \, \mathrm{TeV}}} \, I_{\nu_{\mu}, \, \mathrm{atm}} \, A_\mathrm{eff}(E_\nu) \, V(\delta) \, dE_\nu,
\end{equation}
respectively, where $\Delta \Omega = 2 \pi (1-\cos{\theta_\mathrm{obs}})$. $A_\mathrm{eff}(E_\nu)$ and $V(\delta)$ are the effective area of KM3NeT \citep{km3net_2016JPhG...43h4001A} and the fraction of time in which a source located at $\delta$ is visible below the horizon ($V \sim 0.95$ for Wd 1 \citep{km3net_2019APh...111..100A}). For $T_\mathrm{obs}=1 \, \mathrm{year}$, the estimated $N_\mathrm{s}$ and $N_\mathrm{b}$ are 3.62 and 8.16, which corresponds to $\sigma \sim 1.27$. Therefore, we expect a detection significance of $\sim 3.8\sigma$ for 10 years of observation of Wd 1 with KM3NeT.

\section{Discussion and conclusions}
\label{sec:discuss}

In this work, we presented a comprehensive study of the gamma-ray emission from Wd 1. We calculated the spectral and spatial distributions of protons and electrons assuming that particle acceleration occurs at the collective wind termination shock. The subsequent non-thermal emission was calculated and compared with the multi-wavelength observations by Planck, eROSITA, Fermi-LAT and HESS.

Our results show that in the case of Bohm diffusion, both the leptonic and hadronic scenarios can account for the gamma-ray spectrum and morphology of Wd 1, while not violating the radio and X-ray upper limits. However, in the case of Kraichnan diffusion, only the hadronic scenario can explain the observations, while the leptonic scenario fails to reproduce the highest-energy gamma-ray data points. For the leptonic scenario, we assume that the gas inside the bubble is from the stellar winds, with no pre-existing gas clouds surviving after the passage the forward shock. For a given $K_\mathrm{ep} \sim 0.01$, gamma-ray emission is dominated by the radiation produced by electrons through ICS. The corresponding hadronic gamma-ray emission represents the limiting case, where the target gas density is limited to the one in the shocked wind. The eROSITA X-ray upper limit constrains the magnetic field immediately downstream within $\lesssim$ a few $\mu$G. On the other hand, for the hadronic scenario with $K_\mathrm{ep} \sim 10^{-3}$, the gas density required to reproduce the gamma-ray emission should be $\gtrsim 1.2~\mathrm{cm}^{-3}$. As shown in Section \ref{sec:env:gas}, there is evidence for sufficient gas in the bubble to motivate the production of gamma-ray emission through pp interactions. Although there is no clear correlation between the gamma-ray peaks and gas clouds in the HESS significance maps, this can be interpreted in terms of the effect of the self-absorption of HI \citep{Sano_2026ApJ..1004...12S}. We note that a higher $K_\mathrm{ep}$ can increase the leptonic contribution, but may lead to overshooting the eROSITA upper limit and/or the Fermi-LAT GeV data. Our lepto-hadronic analysis suggests that both hadronic and leptonic processes may play an important role in accounting for the gamma-ray emission of Wd 1. Current multi-wavelength observations are still insufficient to distinguish the effects of these two processes, due to their limited spatial and spectral resolutions. Future deeper X-ray observations with NewAthena, neutrino observations with KM3NeT, and high resolution observations with CTAO in the GeV range, may provide us more information in understanding the origin of the gamma-ray emission toward Wd 1. Better spatial separation of the different gas components is also essential for addressing this problem.

In our modeling, we considered two different diffusion models: Bohm and Kraichnan diffusion. For Bohm diffusion, both leptonic and hadronic scenarios are able to account for the data. On the other hand, for the case of Kraichnan diffusion with a smaller $L_\mathrm{c}\sim0.2~\mathrm{pc}$, the hadronic scenario provides a better fit than the leptonic scenario, as the latter fails to reproduce the highest-energy data points due to more severe energy losses experienced by electrons. This finding is important because, in the hadronic scenario, it is not necessary to invoke Bohm diffusion, which is unlikely to be realized in the environment of a star cluster \citep{Morlino_2021,Blasi_2023}. It provides a clear contrast to the case of Cygnus OB2 association, where Bohm diffusion is required to explain the observed gamma-ray emission \citep{Li_2026}. It is worth noting that the energy of the highest-energy gamma rays from Cygnus OB2 is about one order of magnitude higher than for Wd 1, but it has been suggested that at least part of such emission could be due to the microquasar Cygnus X-3 \cite[]{Li_2026,Shi_cygx3}. We made a quantitative assessment of the role of future observations with neutrino telescopes to clarify the origin of the gamma ray emission from Wd 1. The next generation gamma-ray observatory in the southern hemisphere, SWGO, may also provide us the opportunity to probe the high-energy part of the gamma ray spectrum of Wd 1.

\begin{acknowledgements}
We thank Stefano Menchiari and Giada Peron for valuable discussions on gas data analysis. This work has been partially funded by the European Union - Next Generation EU under the project IR0000012---CTA+ ``Rafforzamento e creazione di IR nell’ambito del Piano Nazionale di Ripresa e Resilienza (PNRR)'', under PRIN-MUR 2022TJW4EJ ``Unveiling the footprints of the cosmic ray journey through the Galaxy and beyond'' and under MUR National Innovation Ecosystem grant ECS00000041 - VITALITY/ASTRA - CUP D13C21000430001. 
\end{acknowledgements}

\bibliographystyle{aa}
\bibliography{bib_ben}

@ARTICLE{Subedi2017,
       author = {{Subedi}, P. and {Sonsrettee}, W. and {Blasi}, P. and {Ruffolo}, D. and {Matthaeus}, W.~H. and {Montgomery}, D. and {Chuychai}, P. and {Dmitruk}, P. and {Wan}, M. and {Parashar}, T.~N. and {Chhiber}, R.},
        title = "{Charged Particle Diffusion in Isotropic Random Magnetic Fields}",
      journal = {\apj},
         year = 2017,
        month = mar,
       volume = {837},
       number = {2},
          eid = {140},
        pages = {140},
          doi = {10.3847/1538-4357/aa603a},
archivePrefix = {arXiv},
       eprint = {1612.09507},
 primaryClass = {physics.space-ph},
       adsurl = {https://ui.adsabs.harvard.edu/abs/2017ApJ...837..140S}
}

@ARTICLE{Cristofari2021,
       author = {{Cristofari}, P. and {Blasi}, P. and {Caprioli}, D.},
        title = "{Cosmic ray protons and electrons from supernova remnants}",
      journal = {\aap},
         year = 2021,
        month = jun,
       volume = {650},
          eid = {A62},
        pages = {A62},
          doi = {10.1051/0004-6361/202140448},
archivePrefix = {arXiv},
       eprint = {2103.02375},
 primaryClass = {astro-ph.HE},
       adsurl = {https://ui.adsabs.harvard.edu/abs/2021A&A...650A..62C}
}

@ARTICLE{Cristofari2020,
       author = {{Cristofari}, Pierre and {Blasi}, Pasquale and {Amato}, Elena},
        title = "{The low rate of Galactic pevatrons}",
      journal = {Astroparticle Physics},
         year = 2020,
        month = dec,
       volume = {123},
          eid = {102492},
        pages = {102492},
          doi = {10.1016/j.astropartphys.2020.102492},
archivePrefix = {arXiv},
       eprint = {2007.04294},
 primaryClass = {astro-ph.HE},
       adsurl = {https://ui.adsabs.harvard.edu/abs/2020APh...12302492C}
}

@ARTICLE{Capanema2026,
       author = {{Capanema}, A. and {Blasi}, P. and {Sobacchi}, E.},
        title = "{Acoustic instability at shockwave precursors}",
      journal = {\aap},
         year = 2026,
        month = may,
       volume = {710},
          eid = {A14},
        pages = {A14},
          doi = {10.1051/0004-6361/202659871},
archivePrefix = {arXiv},
       eprint = {2604.12514},
 primaryClass = {astro-ph.HE},
       adsurl = {https://ui.adsabs.harvard.edu/abs/2026A&A...710A..14C}
}

@ARTICLE{DruryDownes2012,
       author = {{Drury}, L. O'C. and {Downes}, T.~P.},
        title = "{Turbulent magnetic field amplification driven by cosmic ray pressure gradients}",
      journal = {\mnras},
         year = 2012,
        month = dec,
       volume = {427},
       number = {3},
        pages = {2308-2313},
          doi = {10.1111/j.1365-2966.2012.22106.x},
archivePrefix = {arXiv},
       eprint = {1205.6823},
 primaryClass = {astro-ph.HE},
       adsurl = {https://ui.adsabs.harvard.edu/abs/2012MNRAS.427.2308D}
}

@ARTICLE{DruryFalle1986,
       author = {{Drury}, L.~O. and {Falle}, S.~A.~E.~G.},
        title = "{On the Stability of Shocks Modified by Particle Acceleration}",
      journal = {\mnras},
         year = 1986,
        month = nov,
       volume = {223},
        pages = {353},
          doi = {10.1093/mnras/223.2.353},
       adsurl = {https://ui.adsabs.harvard.edu/abs/1986MNRAS.223..353D}
}

@ARTICLE{Jones2009,
       author = {{Beresnyak}, A. and {Jones}, T.~W. and {Lazarian}, A.},
        title = "{Turbulence-Induced Magnetic Fields and Structure of Cosmic Ray Modified Shocks}",
      journal = {\apj},
         year = 2009,
        month = dec,
       volume = {707},
       number = {2},
        pages = {1541-1549},
          doi = {10.1088/0004-637X/707/2/1541},
archivePrefix = {arXiv},
       eprint = {0908.2806},
 primaryClass = {astro-ph.HE},
       adsurl = {https://ui.adsabs.harvard.edu/abs/2009ApJ...707.1541B}
}

@BOOK{Gaisser_2016,
       author = {{Gaisser}, Thomas K. and {Engel}, Ralph and {Resconi}, Elisa},
        title = "{Cosmic Rays and Particle Physics}",
         year = 2016,
       adsurl = {https://ui.adsabs.harvard.edu/abs/2016crpp.book.....G}
}

@ARTICLE{Hillas_1984,
       author = {{Hillas}, A.~M.},
        title = "{The Origin of Ultra-High-Energy Cosmic Rays}",
      journal = {\araa},
         year = 1984,
        month = jan,
       volume = {22},
        pages = {425-444},
          doi = {10.1146/annurev.aa.22.090184.002233},
       adsurl = {https://ui.adsabs.harvard.edu/abs/1984ARA&A..22..425H}
}

@ARTICLE{PortegiesZwart_2010,
       author = {{Portegies Zwart}, Simon F. and {McMillan}, Stephen L.~W. and {Gieles}, Mark},
        title = "{Young Massive Star Clusters}",
      journal = {\araa},
         year = 2010,
        month = sep,
       volume = {48},
        pages = {431-493},
          doi = {10.1146/annurev-astro-081309-130834},
archivePrefix = {arXiv},
       eprint = {1002.1961},
 primaryClass = {astro-ph.GA},
       adsurl = {https://ui.adsabs.harvard.edu/abs/2010ARA&A..48..431P}
}

@ARTICLE{Peron_2025,
       author = {{Peron}, G. and {Menchiari}, S. and {Morlino}, G. and {Amato}, E.},
        title = "{Hadronic acceleration in the young star cluster NGC 6611 inside the M16 region unveiled by Fermi-LAT: Constraints on the acceleration efficiency}",
      journal = {\aap},
         year = 2025,
        month = nov,
       volume = {703},
          eid = {L8},
        pages = {L8},
          doi = {10.1051/0004-6361/202556564},
archivePrefix = {arXiv},
       eprint = {2507.13009},
 primaryClass = {astro-ph.HE},
       adsurl = {https://ui.adsabs.harvard.edu/abs/2025A&A...703L...8P}
}

@ARTICLE{Li_2026,
       author = {{Li}, Ben and {Blasi}, Pasquale and {Amato}, Elena},
        title = "{Suppressed diffusion and gamma-ray emission from the Cygnus bubble}",
      journal = {\aap},
         year = 2026,
        month = aug,
       volume = {712},
          eid = {A41},
        pages = {A41},
          doi = {10.1051/0004-6361/202558034},
archivePrefix = {arXiv},
       eprint = {2606.03881},
 primaryClass = {astro-ph.HE},
       adsurl = {https://ui.adsabs.harvard.edu/abs/2026A&A...712A..41L}
}

@ARTICLE{Harer_2025,
       author = {{H{\"a}rer}, L. and {Vieu}, T. and {Schulze}, F. and {Larkin}, C.~J.~K. and {Reville}, B.},
        title = "{Deciphering the gamma-ray emission in the Cygnus region}",
      journal = {\aap},
         year = 2025,
        month = nov,
       volume = {703},
          eid = {A111},
        pages = {A111},
          doi = {10.1051/0004-6361/202555531},
archivePrefix = {arXiv},
       eprint = {2508.21644},
 primaryClass = {astro-ph.HE},
       adsurl = {https://ui.adsabs.harvard.edu/abs/2025A&A...703A.111H}
}

@ARTICLE{Westerlund2022,
       author = {{Aharonian}, F. and {Ashkar}, H. and {Backes}, M. and {Barbosa Martins}, V. and {Becherini}, Y. and {Berge}, D. and {Bi}, B. and {B{\"o}ttcher}, M. and {de Bony de Lavergne}, M. and {Bradascio}, F. and {Brose}, R. and {Brun}, F. and {Bulik}, T. and {Burger-Scheidlin}, C. and {Cangemi}, F. and {Caroff}, S. and {Casanova}, S. and {Cerruti}, M. and {Chand}, T. and {Chandra}, S. and {Chen}, A. and {Chibueze}, O. and {Cristofari}, P. and {Damascene Mbarubucyeye}, J. and {Djannati-Ata{\"\i}}, A. and {Ernenwein}, J. -P. and {Feijen}, K. and {Fichet de Clairfontaine}, G. and {Fontaine}, G. and {Funk}, S. and {Gabici}, S. and {Gallant}, Y.~A. and {Ghafourizadeh}, S. and {Giavitto}, G. and {Giunti}, L. and {Glawion}, D. and {Glicenstein}, J.~F. and {Goswami}, P. and {Grondin}, M. -H. and {H{\"a}rer}, L.~K. and {Haupt}, M. and {Hinton}, J.~A. and {H{\"o}rbe}, M. and {Hofmann}, W. and {Holch}, T.~L. and {Holler}, M. and {Horns}, D. and {Jamrozy}, M. and {Joshi}, V. and {Jung-Richardt}, I. and {Kasai}, E. and {Katarzy{\'n}ski}, K. and {Katz}, U. and {Kh{\'e}lifi}, B. and {Klu{\'z}niak}, W. and {Komin}, Nu. and {Kosack}, K. and {Kostunin}, D. and {Kukec Mezek}, G. and {Lang}, R.~G. and {Le Stum}, S. and {Lemi{\`e}re}, A. and {Lemoine-Goumard}, M. and {Lenain}, J. -P. and {Leuschner}, F. and {Lohse}, T. and {Luashvili}, A. and {Lypova}, I. and {Mackey}, J. and {Majumdar}, J. and {Malyshev}, D. and {Marandon}, V. and {Marchegiani}, P. and {Marcowith}, A. and {Mart{\'\i}-Devesa}, G. and {Marx}, R. and {Maurin}, G. and {Meyer}, M. and {Mitchell}, A. and {Moderski}, R. and {Mohrmann}, L. and {Montanari}, A. and {Moulin}, E. and {Muller}, J. and {Murach}, T. and {Nakashima}, K. and {de Naurois}, M. and {Nayerhoda}, A. and {Niemiec}, J. and {Ohm}, S. and {Olivera-Nieto}, L. and {de Ona Wilhelmi}, E. and {Ostrowski}, M. and {Panny}, S. and {Panter}, M. and {Parsons}, R.~D. and {Peron}, G. and {Prokhorov}, D.~A. and {P{\"u}hlhofer}, G. and {Punch}, M. and {Quirrenbach}, A. and {Rauth}, R. and {Reichherzer}, P. and {Reimer}, A. and {Reimer}, O. and {Renaud}, M. and {Reville}, B. and {Rieger}, F. and {Rowell}, G. and {Rudak}, B. and {Ruiz-Velasco}, E. and {Sahakian}, V. and {Salzmann}, H. and {Sanchez}, D.~A. and {Santangelo}, A. and {Sasaki}, M. and {Sch{\"u}ssler}, F. and {Schutte}, H.~M. and {Schwanke}, U. and {Shapopi}, J.~N.~S. and {Specovius}, A. and {Spencer}, S. and {Stawarz}, {\L}. and {Steenkamp}, R. and {Steinmassl}, S. and {Steppa}, C. and {Sushch}, I. and {Suzuki}, H. and {Takahashi}, T. and {Tanaka}, T. and {Terrier}, R. and {Thorpe-Morgan}, C. and {Tsirou}, M. and {Tsuji}, N. and {Tuffs}, R. and {Unbehaun}, T. and {van Eldik}, C. and {van Soelen}, B. and {Vecchi}, M. and {Veh}, J. and {Venter}, C. and {Vink}, J. and {Wagner}, S.~J. and {White}, R. and {Wierzcholska}, A. and {Wong}, Y. Wun and {Zacharias}, M. and {Zargaryan}, D. and {Zdziarski}, A.~A. and {Zhu}, S.~J. and {Zouari}, S. and {{\.Z}ywucka}, N. and {Blackwell}, R. and {Braiding}, C. and {Burton}, M. and {Cubuk}, K. and {Filipovi{\'c}}, M. and {Tothill}, N. and {Wong}, G.},
        title = "{A deep spectromorphological study of the {\ensuremath{\gamma}}-ray emission surrounding the young massive stellar cluster Westerlund 1}",
      journal = {\aap},
         year = 2022,
        month = oct,
       volume = {666},
          eid = {A124},
        pages = {A124},
          doi = {10.1051/0004-6361/202244323},
archivePrefix = {arXiv},
       eprint = {2207.10921},
 primaryClass = {astro-ph.HE},
       adsurl = {https://ui.adsabs.harvard.edu/abs/2022A&A...666A.124A}
}

@ARTICLE{Harer_2023,
       author = {{H{\"a}rer}, Lucia K. and {Reville}, Brian and {Hinton}, Jim and {Mohrmann}, Lars and {Vieu}, Thibault},
        title = "{Understanding the TeV {\ensuremath{\gamma}}-ray emission surrounding the young massive star cluster Westerlund 1}",
      journal = {\aap},
         year = 2023,
        month = mar,
       volume = {671},
          eid = {A4},
        pages = {A4},
          doi = {10.1051/0004-6361/202245444},
archivePrefix = {arXiv},
       eprint = {2301.10496},
 primaryClass = {astro-ph.HE},
       adsurl = {https://ui.adsabs.harvard.edu/abs/2023A&A...671A...4H}
}

@ARTICLE{Lemoine-Goumard_2025NatCo..1610820L,
       author = {{Lemoine-Goumard}, Marianne and {H{\"a}rer}, Lucia and {Mohrmann}, Lars and {Bernet}, Romain and {Hinton}, Jim and {Peron}, Giada and {Reville}, Brian and {Tibaldo}, Luigi and {Vieu}, Thibault},
        title = "{A cosmic-ray loaded nascent outflow driven by a massive star cluster}",
      journal = {Nature Communications},
         year = 2025,
        month = dec,
       volume = {16},
       number = {1},
          eid = {10820},
        pages = {10820},
          doi = {10.1038/s41467-025-65592-4},
       adsurl = {https://ui.adsabs.harvard.edu/abs/2025NatCo..1610820L}
}

@ARTICLE{Shi_2026JHEAp..5100560S,
       author = {{Shi}, Zhaodong and {Yang}, Rui-Zhi},
        title = "{On the hadronic origin of the very high energy {\ensuremath{\gamma}}-ray emission surrounding the young massive stellar cluster Westerlund 1}",
      journal = {Journal of High Energy Astrophysics},
         year = 2026,
        month = mar,
       volume = {51},
          eid = {100560},
        pages = {100560},
          doi = {10.1016/j.jheap.2026.100560},
archivePrefix = {arXiv},
       eprint = {2601.17324},
 primaryClass = {astro-ph.HE},
       adsurl = {https://ui.adsabs.harvard.edu/abs/2026JHEAp..5100560S}
}

@ARTICLE{Sano_2026ApJ..1004...12S,
       author = {{Sano}, H. and {Fukui}, Y. and {Fujimori}, S. and {Murase}, T. and {Alsaberi}, R.~Z.~E. and {Filipovi{\'c}}, M.~D. and {Rowell}, G. and {Aruga}, M. and {Asano}, Y. and {Bhuvana}, R.~G. and {Demachi}, F. and {Einecke}, S. and {Fukaya}, N. and {Hamada}, R. and {Inoue}, H. and {Kamazaki}, T. and {Lazarevi{\'c}}, S. and {Minamidani}, T. and {Smeaton}, Z.~J. and {Sudou}, H. and {Tachihara}, K. and {Takaba}, H. and {Tsuge}, K. and {Yamada}, R.~I.},
        title = "{Discovery of Molecular and Atomic Gas Associated with HESS J1646{\ensuremath{-}}458 (Westerlund 1): Spatial TeV Gamma-Ray and Interstellar Proton Correspondence}",
      journal = {\apj},
         year = 2026,
        month = jun,
       volume = {1004},
       number = {1},
          eid = {12},
        pages = {12},
          doi = {10.3847/1538-4357/ae6337},
archivePrefix = {arXiv},
       eprint = {2605.25467},
 primaryClass = {astro-ph.HE},
       adsurl = {https://ui.adsabs.harvard.edu/abs/2026ApJ..1004...12S}
}

@ARTICLE{Shi_cygx3,
       author = {{Shi}, Zhaodong and {Wang}, Guangwei and {Yang}, Ruizhi and {Aharonian}, Felix},
        title = "{Microquasar Cygnus X-3 as the PeVatron powering the Cygnus Bubble}",
      journal = {arXiv e-prints},
         year = 2026,
        month = jul,
          eid = {arXiv:2607.07100},
        pages = {arXiv:2607.07100},
          doi = {10.48550/arXiv.2607.07100},
archivePrefix = {arXiv},
       eprint = {2607.07100},
 primaryClass = {astro-ph.HE},
       adsurl = {https://ui.adsabs.harvard.edu/abs/2026arXiv260707100S}
}

@ARTICLE{LHAASO_cygx3,
       author = {{The LHAASO Collaboration} and {Cao}, Zhen and {Aharonian}, F. and {Bai}, Y.~X. and {Bao}, Y.~W. and {Bastieri}, D. and {Bi}, X.~J. and {Bi}, Y.~J. and {Bian}, W. and {Blunier}, J. and {Bukevich}, A.~V. and {Cai}, C.~M. and {Cai}, Y.~Y. and {Cao}, W.~Y. and {Cao}, Zhe and {Chang}, J. and {Chang}, J.~F. and {Chen}, E.~S. and {Chen}, G.~H. and {Chen}, H.~K. and {Chen}, L.~F. and {Chen}, Liang and {Chen}, Long and {Chen}, M.~J. and {Chen}, M.~L. and {Chen}, Q.~H. and {Chen}, S. and {Chen}, S.~H. and {Chen}, S.~Z. and {Chen}, T.~L. and {Chen}, X.~B. and {Chen}, X.~J. and {Chen}, X.~P. and {Chen}, Y. and {Cheng}, N. and {Cheng}, Q.~Y. and {Cheng}, Y.~D. and {Cui}, M.~Y. and {Cui}, S.~W. and {Cui}, X.~H. and {Cui}, Y.~D. and {Dai}, B.~Z. and {Dai}, H.~L. and {Dai}, Z.~G. and {Danzengluobu} and {Diao}, Y.~X. and {Dong}, A.~J. and {Dong}, X.~Q. and {Duan}, K.~K. and {Fan}, J.~H. and {Fan}, Y.~Z. and {Fang}, J. and {Fang}, J.~H. and {Fang}, K. and {Feng}, C.~F. and {Feng}, H. and {Feng}, L. and {Feng}, S.~H. and {Feng}, X.~T. and {Feng}, Y. and {Feng}, Y.~L. and {Gabici}, S. and {Gao}, B. and {Gao}, Q. and {Gao}, W. and {Gao}, W.~K. and {Ge}, M.~M. and {Ge}, T.~T. and {Geng}, L.~S. and {Giacinti}, G. and {Gong}, G.~H. and {Gou}, Q.~B. and {Gu}, M.~H. and {Guo}, F.~L. and {Guo}, J. and {Guo}, K.~J. and {Guo}, X.~L. and {Guo}, Y.~Q. and {Guo}, Y.~Y. and {Han}, R.~P. and {Hannuksela}, O.~A. and {Hasan}, M. and {He}, H.~H. and {He}, H.~N. and {He}, J.~Y. and {He}, X.~Y. and {He}, Y. and {Hern{\'a}ndez-Cadena}, S. and {Hou}, B.~W. and {Hou}, C. and {Hou}, X. and {Hu}, H.~B. and {Hu}, S.~C. and {Huang}, C. and {Huang}, D.~H. and {Huang}, J.~J. and {Huang}, X.~L. and {Huang}, X.~T. and {Huang}, X.~Y. and {Huang}, Y. and {Huang}, Y.~Y. and {Inventar}, A. and {Ji}, X.~L. and {Jia}, H.~Y. and {Jia}, K. and {Jiang}, H.~B. and {Jiang}, K. and {Jiang}, X.~W. and {Jiang}, Z.~J. and {Jin}, M. and {Kaci}, S. and {Kang}, M.~M. and {Karpikov}, I. and {Khangulyan}, D. and {Kuleshov}, D. and {Kurinov}, K. and {Li}, Cheng and {Li}, Cong and {Li}, D. and {Li}, F. and {Li}, H.~B. and {Li}, H.~C. and {Li}, Jian and {Li}, Jie and {Li}, K. and {Li}, L. and {Li}, R.~L. and {Li}, S.~D. and {Li}, T.~Y. and {Li}, W.~L. and {Li}, X.~R. and {Li}, Xin and {Li}, Y. and {Li}, Zhe and {Li}, Zhuo and {Liang}, E.~W. and {Liang}, Y.~F. and {Lin}, S.~J. and {Liu}, B. and {Liu}, C. and {Liu}, D. and {Liu}, D.~B. and {Liu}, H. and {Liu}, J. and {Liu}, J.~L. and {Liu}, J.~R. and {Liu}, M.~Y. and {Liu}, R.~Y. and {Liu}, S.~M. and {Liu}, W. and {Liu}, X. and {Liu}, Y. and {Liu}, Y. and {Liu}, Y.~N. and {Lou}, Y.~Q. and {Luo}, Q. and {Luo}, Y. and {Lv}, H.~K. and {Ma}, B.~Q. and {Ma}, L.~L. and {Ma}, X.~H. and {Maliy}, I.~O. and {Mao}, J.~R. and {Min}, Z. and {Mitthumsiri}, W. and {Mizuno}, Y. and {Mou}, G.~B. and {Neronov}, A. and {Ng}, K.~C.~Y. and {Ni}, M.~Y. and {Nie}, L. and {Ou}, L.~J. and {Ou}, Z.~W. and {Pattarakijwanich}, P. and {Pei}, Z.~Y. and {Peng}, D.~Y. and {Qi}, J.~C. and {Qi}, M.~Y. and {Qin}, J.~J. and {Qu}, D. and {Raza}, A. and {Ren}, C.~Y. and {Ruffolo}, D. and {S{\'a}iz}, A. and {Savchenko}, D. and {Semikoz}, D. and {Shao}, L. and {Shchegolev}, O. and {Shen}, Y.~Z. and {Sheng}, X.~D. and {Shi}, Z.~D. and {Shu}, F.~W. and {Song}, H.~C. and {Stenkin}, Yu. V. and {Stepanov}, V. and {Su}, Y. and {Sun}, D.~X. and {Sun}, H. and {Sun}, J.~X. and {Sun}, Q.~N.},
        title = "{Cygnus X-3: A variable petaelectronvolt gamma-ray source}",
      journal = {arXiv e-prints},
         year = 2025,
        month = dec,
          eid = {arXiv:2512.16638},
        pages = {arXiv:2512.16638},
          doi = {10.48550/arXiv.2512.16638},
archivePrefix = {arXiv},
       eprint = {2512.16638},
 primaryClass = {astro-ph.HE},
       adsurl = {https://ui.adsabs.harvard.edu/abs/2025arXiv251216638T}
}

@ARTICLE{Bell_2004,
       author = {{Bell}, A.~R.},
        title = "{Turbulent amplification of magnetic field and diffusive shock acceleration of cosmic rays}",
      journal = {\mnras},
         year = 2004,
        month = sep,
       volume = {353},
       number = {2},
        pages = {550-558},
          doi = {10.1111/j.1365-2966.2004.08097.x},
       adsurl = {https://ui.adsabs.harvard.edu/abs/2004MNRAS.353..550B}
}

@ARTICLE{Amato_2009,
       author = {{Amato}, E. and {Blasi}, P.},
        title = "{A kinetic approach to cosmic-ray-induced streaming instability at supernova shocks}",
      journal = {\mnras},
         year = 2009,
        month = feb,
       volume = {392},
       number = {4},
        pages = {1591-1600},
          doi = {10.1111/j.1365-2966.2008.14200.x},
archivePrefix = {arXiv},
       eprint = {0806.1223},
 primaryClass = {astro-ph},
       adsurl = {https://ui.adsabs.harvard.edu/abs/2009MNRAS.392.1591A}
}

@ARTICLE{Bell_1978,
       author = {{Bell}, A.~R.},
        title = "{The acceleration of cosmic rays in shock fronts - I.}",
      journal = {\mnras},
         year = 1978,
        month = jan,
       volume = {182},
        pages = {147-156},
          doi = {10.1093/mnras/182.2.147},
       adsurl = {https://ui.adsabs.harvard.edu/abs/1978MNRAS.182..147B}
}

@ARTICLE{Blandford_1978,
       author = {{Blandford}, R.~D. and {Ostriker}, J.~P.},
        title = "{Particle acceleration by astrophysical shocks.}",
      journal = {\apjl},
         year = 1978,
        month = apr,
       volume = {221},
        pages = {L29-L32},
          doi = {10.1086/182658},
       adsurl = {https://ui.adsabs.harvard.edu/abs/1978ApJ...221L..29B}
}

@article{Drury_1983,
        doi = {10.1088/0034-4885/46/8/002},
        url = {https://doi.org/10.1088/0034-4885/46/8/002},
        year = {1983},
        month = {aug},
        publisher = {},
        volume = {46},
        number = {8},
        pages = {973},
        author = {L O'C Drury},
        title = {An introduction to the theory of diffusive shock acceleration of energetic particles in tenuous plasmas},
        journal = {Reports on Progress in Physics}
}

@ARTICLE{Malkov_2001,
       author = {{Malkov}, M.~A. and {Drury}, L. O'C.},
        title = "{Nonlinear theory of diffusive acceleration of particles by shock waves}",
      journal = {Reports on Progress in Physics},
         year = 2001,
        month = apr,
       volume = {64},
       number = {4},
        pages = {429-481},
          doi = {10.1088/0034-4885/64/4/201},
       adsurl = {https://ui.adsabs.harvard.edu/abs/2001RPPh...64..429M}
}

@ARTICLE{Lagage_1983,
       author = {{Lagage}, P.~O. and {Cesarsky}, C.~J.},
        title = "{The maximum energy of cosmic rays accelerated by supernova shocks.}",
      journal = {\aap},
         year = 1983,
        month = sep,
       volume = {125},
        pages = {249-257},
       adsurl = {https://ui.adsabs.harvard.edu/abs/1983A&A...125..249L}
}

@ARTICLE{Bell_2013,
       author = {{Bell}, A.~R. and {Schure}, K.~M. and {Reville}, B. and {Giacinti}, G.},
        title = "{Cosmic-ray acceleration and escape from supernova remnants}",
      journal = {\mnras},
         year = 2013,
        month = may,
       volume = {431},
       number = {1},
        pages = {415-429},
          doi = {10.1093/mnras/stt179},
archivePrefix = {arXiv},
       eprint = {1301.7264},
 primaryClass = {astro-ph.HE},
       adsurl = {https://ui.adsabs.harvard.edu/abs/2013MNRAS.431..415B}
}

@ARTICLE{Schure_2013,
       author = {{Schure}, K.~M. and {Bell}, A.~R.},
        title = "{Cosmic ray acceleration in young supernova remnants}",
      journal = {\mnras},
         year = 2013,
        month = oct,
       volume = {435},
       number = {2},
        pages = {1174-1185},
          doi = {10.1093/mnras/stt1371},
archivePrefix = {arXiv},
       eprint = {1307.6575},
 primaryClass = {astro-ph.HE},
       adsurl = {https://ui.adsabs.harvard.edu/abs/2013MNRAS.435.1174S}
}

@article{Klepach_2000,
        title = {Cosmic ray acceleration by multiple spherical shocks},
        journal = {Astroparticle Physics},
        volume = {13},
        number = {2},
        pages = {161-172},
        year = {2000},
        issn = {0927-6505},
        doi = {https://doi.org/10.1016/S0927-6505(99)00108-5},
        url = {https://www.sciencedirect.com/science/article/pii/S0927650599001085},
        author = {E.G. Klepach and V.S. Ptuskin and V.N. Zirakashvili}
}

@ARTICLE{Bykov&Toptygin_2001,
       author = {{Bykov}, A.~M. and {Toptygin}, I.~N.},
        title = "{A Model of Particle Acceleration to High Energies by Multiple Supernova Explosions in OB Associations}",
      journal = {Astronomy Letters},
         year = 2001,
        month = oct,
       volume = {27},
       number = {10},
        pages = {625-633},
          doi = {10.1134/1.1404456},
       adsurl = {https://ui.adsabs.harvard.edu/abs/2001AstL...27..625B}
}

@ARTICLE{Vieu+2022,
       author = {{Vieu}, T. and {Gabici}, S. and {Tatischeff}, V. and {Ravikularaman}, S.},
        title = "{Cosmic ray production in superbubbles}",
      journal = {\mnras},
         year = 2022,
        month = may,
       volume = {512},
       number = {1},
        pages = {1275-1293},
          doi = {10.1093/mnras/stac543},
archivePrefix = {arXiv},
       eprint = {2201.07488},
 primaryClass = {astro-ph.HE},
       adsurl = {https://ui.adsabs.harvard.edu/abs/2022MNRAS.512.1275V}
}

@article{Blasi_2023,
    author = {Blasi, Pasquale and Morlino, Giovanni},
    title = {High-energy cosmic rays and gamma-rays from star clusters: the case of Cygnus OB2},
    journal = {Monthly Notices of the Royal Astronomical Society},
    volume = {523},
    number = {3},
    pages = {4015-4028},
    year = {2023},
    month = {06},
    issn = {0035-8711},
    doi = {10.1093/mnras/stad1662},
    url = {https://doi.org/10.1093/mnras/stad1662},
    eprint = {https://academic.oup.com/mnras/article-pdf/523/3/4015/50617929/stad1662.pdf},
}

@ARTICLE{Morlino_2021,
       author = {{Morlino}, G. and {Blasi}, P. and {Peretti}, E. and {Cristofari}, P.},
        title = "{Particle acceleration in winds of star clusters}",
      journal = {\mnras},
         year = 2021,
        month = jul,
       volume = {504},
       number = {4},
        pages = {6096-6105},
          doi = {10.1093/mnras/stab690},
archivePrefix = {arXiv},
       eprint = {2102.09217},
 primaryClass = {astro-ph.HE},
       adsurl = {https://ui.adsabs.harvard.edu/abs/2021MNRAS.504.6096M}
}

@ARTICLE{Menchiari_2023,
       author = {{Menchiari}, Stefano},
        title = "{Probing star clusters as cosmic ray factories}",
      journal = {arXiv e-prints},
         year = 2023,
        month = jul,
          eid = {arXiv:2307.03477},
        pages = {arXiv:2307.03477},
          doi = {10.48550/arXiv.2307.03477},
archivePrefix = {arXiv},
       eprint = {2307.03477},
 primaryClass = {astro-ph.HE},
       adsurl = {https://ui.adsabs.harvard.edu/abs/2023arXiv230703477M}
}

@ARTICLE{Strong_2007,
       author = {{Strong}, Andrew W. and {Moskalenko}, Igor V. and {Ptuskin}, Vladimir S.},
        title = "{Cosmic-Ray Propagation and Interactions in the Galaxy}",
      journal = {Annual Review of Nuclear and Particle Science},
         year = 2007,
        month = nov,
       volume = {57},
       number = {1},
        pages = {285-327},
          doi = {10.1146/annurev.nucl.57.090506.123011},
archivePrefix = {arXiv},
       eprint = {astro-ph/0701517},
 primaryClass = {astro-ph},
       adsurl = {https://ui.adsabs.harvard.edu/abs/2007ARNPS..57..285S}
}

@ARTICLE{Castor_1975,
       author = {{Castor}, J. and {McCray}, R. and {Weaver}, R.},
        title = "{Interstellar bubbles.}",
      journal = {\apjl},
         year = 1975,
        month = sep,
       volume = {200},
        pages = {L107-L110},
          doi = {10.1086/181908},
       adsurl = {https://ui.adsabs.harvard.edu/abs/1975ApJ...200L.107C}
}

@ARTICLE{Weaver_1977,
       author = {{Weaver}, R. and {McCray}, R. and {Castor}, J. and {Shapiro}, P. and {Moore}, R.},
        title = "{Interstellar bubbles. II. Structure and evolution.}",
      journal = {\apj},
         year = 1977,
        month = dec,
       volume = {218},
        pages = {377-395},
          doi = {10.1086/155692},
       adsurl = {https://ui.adsabs.harvard.edu/abs/1977ApJ...218..377W}
}

@ARTICLE{Gupta_2018,
       author = {{Gupta}, Siddhartha and {Nath}, Biman B. and {Sharma}, Prateek and {Eichler}, David},
        title = "{Lack of thermal energy in superbubbles: hint of cosmic rays?}",
      journal = {\mnras},
         year = 2018,
        month = jan,
       volume = {473},
       number = {2},
        pages = {1537-1553},
          doi = {10.1093/mnras/stx2427},
archivePrefix = {arXiv},
       eprint = {1705.10448},
 primaryClass = {astro-ph.GA},
       adsurl = {https://ui.adsabs.harvard.edu/abs/2018MNRAS.473.1537G}
}

@ARTICLE{Kelner_2006,
       author = {{Kelner}, S.~R. and {Aharonian}, F.~A. and {Bugayov}, V.~V.},
        title = "{Energy spectra of gamma rays, electrons, and neutrinos produced at proton-proton interactions in the very high energy regime}",
      journal = {\prd},
         year = 2006,
        month = aug,
       volume = {74},
       number = {3},
          eid = {034018},
        pages = {034018},
          doi = {10.1103/PhysRevD.74.034018},
archivePrefix = {arXiv},
       eprint = {astro-ph/0606058},
 primaryClass = {astro-ph},
       adsurl = {https://ui.adsabs.harvard.edu/abs/2006PhRvD..74c4018K}
}

@ARTICLE{Kafexhiu_2014,
       author = {{Kafexhiu}, Ervin and {Aharonian}, Felix and {Taylor}, Andrew M. and {Vila}, Gabriela S.},
        title = "{Parametrization of gamma-ray production cross sections for p p interactions in a broad proton energy range from the kinematic threshold to PeV energies}",
      journal = {\prd},
         year = 2014,
        month = dec,
       volume = {90},
       number = {12},
          eid = {123014},
        pages = {123014},
          doi = {10.1103/PhysRevD.90.123014},
archivePrefix = {arXiv},
       eprint = {1406.7369},
 primaryClass = {astro-ph.HE},
       adsurl = {https://ui.adsabs.harvard.edu/abs/2014PhRvD..90l3014K}
}

@ARTICLE{Krakau_2015,
       author = {{Krakau}, S. and {Schlickeiser}, R.},
        title = "{Pion Production Momentum Loss of Cosmic Ray Hadrons}",
      journal = {\apj},
         year = 2015,
        month = apr,
       volume = {802},
       number = {2},
          eid = {114},
        pages = {114},
          doi = {10.1088/0004-637X/802/2/114},
       adsurl = {https://ui.adsabs.harvard.edu/abs/2015ApJ...802..114K}
}

@ARTICLE{Aharonian_2019,
       author = {{Aharonian}, Felix and {Yang}, Ruizhi and {de O{\~n}a Wilhelmi}, Emma},
        title = "{Massive stars as major factories of Galactic cosmic rays}",
      journal = {Nature Astronomy},
         year = 2019,
        month = mar,
       volume = {3},
        pages = {561-567},
          doi = {10.1038/s41550-019-0724-0},
archivePrefix = {arXiv},
       eprint = {1804.02331},
 primaryClass = {astro-ph.HE},
       adsurl = {https://ui.adsabs.harvard.edu/abs/2019NatAs...3..561A}
}

@ARTICLE{LHAASO_2024,
       author = {{Lhaaso Collaboration}},
        title = "{An ultrahigh-energy {\ensuremath{\gamma}} -ray bubble powered by a super PeVatron}",
      journal = {Science Bulletin},
         year = 2024,
        month = feb,
       volume = {69},
       number = {4},
        pages = {449-457},
          doi = {10.1016/j.scib.2023.12.040},
archivePrefix = {arXiv},
       eprint = {2310.10100},
 primaryClass = {astro-ph.HE},
       adsurl = {https://ui.adsabs.harvard.edu/abs/2024SciBu..69..449L}
}

@ARTICLE{GASS_2009,
       author = {{McClure-Griffiths}, N.~M. and {Pisano}, D.~J. and {Calabretta}, M.~R. and {Ford}, H. Alyson and {Lockman}, Felix J. and {Staveley-Smith}, L. and {Kalberla}, P.~M.~W. and {Bailin}, J. and {Dedes}, L. and {Janowiecki}, S. and {Gibson}, B.~K. and {Murphy}, T. and {Nakanishi}, H. and {Newton-McGee}, K.},
        title = "{Gass: The Parkes Galactic All-Sky Survey. I. Survey Description, Goals, and Initial Data Release}",
      journal = {\apjs},
         year = 2009,
        month = apr,
       volume = {181},
       number = {2},
        pages = {398-412},
          doi = {10.1088/0067-0049/181/2/398},
archivePrefix = {arXiv},
       eprint = {0901.1159},
 primaryClass = {astro-ph.GA},
       adsurl = {https://ui.adsabs.harvard.edu/abs/2009ApJS..181..398M}
}

@ARTICLE{GASS_2015,
       author = {{Kalberla}, P.~M.~W. and {Haud}, U.},
        title = "{GASS: The Parkes Galactic All-Sky Survey. Update: improved correction for instrumental effects and new data release}",
      journal = {\aap},
         year = 2015,
        month = jun,
       volume = {578},
          eid = {A78},
        pages = {A78},
          doi = {10.1051/0004-6361/201525859},
archivePrefix = {arXiv},
       eprint = {1505.01011},
 primaryClass = {astro-ph.GA},
       adsurl = {https://ui.adsabs.harvard.edu/abs/2015A&A...578A..78K}
}

@ARTICLE{Mopra_2018,
       author = {{Braiding}, Catherine and {Wong}, G.~F. and {Maxted}, N.~I. and {Romano}, D. and {Burton}, M.~G. and {Blackwell}, R. and {Filipovi{\'c}}, M.~D. and {Freeman}, M.~S.~R. and {Indermuehle}, B. and {Lau}, J. and {Rebolledo}, D. and {Rowell}, G. and {Snoswell}, C. and {Tothill}, N.~F.~H. and {Voisin}, F. and {de Wilt}, P.},
        title = "{The Mopra Southern Galactic Plane CO Survey{\textemdash}Data Release 3}",
      journal = {\pasa},
         year = 2018,
        month = aug,
       volume = {35},
          eid = {e029},
        pages = {e029},
          doi = {10.1017/pasa.2018.18},
archivePrefix = {arXiv},
       eprint = {1902.04249},
 primaryClass = {astro-ph.GA},
       adsurl = {https://ui.adsabs.harvard.edu/abs/2018PASA...35...29B}
}

@ARTICLE{Mopra_2023,
       author = {{Cubuk}, K.~O. and {Burton}, M.~G. and {Braiding}, C. and {Wong}, G.~F. and {Rowell}, G. and {Maxted}, N.~I. and {Eden}, D. and {Alsaberi}, R.~Z.~E. and {Blackwell}, R. and {Enokiya}, R. and {Feijen}, K. and {Filipovi{\'c}}, M.~D. and {Freeman}, M.~S.~R. and {Fujita}, S. and {Ghavam}, M. and {Gunay}, B. and {Indermuehle}, B. and {Hayashi}, K. and {Kohno}, M. and {Nagaya}, T. and {Nishimura}, A. and {Okawa}, K. and {Rebolledo}, D. and {Romano}, D. and {Sano}, H. and {Snoswell}, C. and {Tothill}, N.~F.~H. and {Tsuge}, K. and {Voisin}, F. and {Yamane}, Y. and {Yoshiike}, S.},
        title = "{The Mopra Southern Galactic Plane CO Survey - data release 4- complete survey}",
      journal = {\pasa},
         year = 2023,
        month = sep,
       volume = {40},
          eid = {e047},
        pages = {e047},
          doi = {10.1017/pasa.2023.44},
       adsurl = {https://ui.adsabs.harvard.edu/abs/2023PASA...40...47C}
}

@ARTICLE{ZOA_1998,
       author = {{Staveley-Smith}, L. and {Juraszek}, S. and {Koribalski}, B.~S. and {Ekers}, R.~D. and {Green}, A.~J. and {Haynes}, R.~F. and {Henning}, P.~A. and {Kesteven}, M.~J. and {Kraan-Korteweg}, R.~C. and {Price}, R.~M. and {Sadler}, E.~M.},
        title = "{New H I-detected Galaxies in the Zone of Avoidance}",
      journal = {\aj},
         year = 1998,
        month = dec,
       volume = {116},
       number = {6},
        pages = {2717-2727},
          doi = {10.1086/300633},
archivePrefix = {arXiv},
       eprint = {astro-ph/9808164},
 primaryClass = {astro-ph},
       adsurl = {https://ui.adsabs.harvard.edu/abs/1998AJ....116.2717S}
}

@ARTICLE{HIPASS_1996,
       author = {{Staveley-Smith}, L. and {Wilson}, W.~E. and {Bird}, T.~S. and {Disney}, M.~J. and {Ekers}, R.~D. and {Freeman}, K.~C. and {Haynes}, R.~F. and {Sinclair}, M.~W. and {Vaile}, R.~A. and {Webster}, R.~L. and {Wright}, A.~E.},
        title = "{The Parkes 21 CM multibeam receiver}",
      journal = {\pasa},
         year = 1996,
        month = nov,
       volume = {13},
       number = {3},
        pages = {243-248},
          doi = {10.1017/S1323358000020919},
       adsurl = {https://ui.adsabs.harvard.edu/abs/1996PASA...13..243S}
}

@ARTICLE{HIPASS_2015,
       author = {{Alves}, Marta I.~R. and {Calabretta}, Mark and {Davies}, Rodney D. and {Dickinson}, Clive and {Staveley-Smith}, Lister and {Davis}, Richard J. and {Chen}, Tianyue and {Barr}, Adam},
        title = "{The HIPASS survey of the Galactic plane in radio recombination lines}",
      journal = {\mnras},
         year = 2015,
        month = jun,
       volume = {450},
       number = {2},
        pages = {2025-2042},
          doi = {10.1093/mnras/stv751},
archivePrefix = {arXiv},
       eprint = {1411.4497},
 primaryClass = {astro-ph.GA},
       adsurl = {https://ui.adsabs.harvard.edu/abs/2015MNRAS.450.2025A}
}

@ARTICLE{Planck_2016,
       author = {{Planck Collaboration} and {Adam}, R. and {Ade}, P.~A.~R. and {Aghanim}, N. and {Alves}, M.~I.~R. and {Arnaud}, M. and {Ashdown}, M. and {Aumont}, J. and {Baccigalupi}, C. and {Banday}, A.~J. and {Barreiro}, R.~B. and {Bartlett}, J.~G. and {Bartolo}, N. and {Battaner}, E. and {Benabed}, K. and {Beno{\^\i}t}, A. and {Benoit-L{\'e}vy}, A. and {Bernard}, J.-P. and {Bersanelli}, M. and {Bielewicz}, P. and {Bock}, J.~J. and {Bonaldi}, A. and {Bonavera}, L. and {Bond}, J.~R. and {Borrill}, J. and {Bouchet}, F.~R. and {Boulanger}, F. and {Bucher}, M. and {Burigana}, C. and {Butler}, R.~C. and {Calabrese}, E. and {Cardoso}, J.-F. and {Catalano}, A. and {Challinor}, A. and {Chamballu}, A. and {Chary}, R.-R. and {Chiang}, H.~C. and {Christensen}, P.~R. and {Clements}, D.~L. and {Colombi}, S. and {Colombo}, L.~P.~L. and {Combet}, C. and {Couchot}, F. and {Coulais}, A. and {Crill}, B.~P. and {Curto}, A. and {Cuttaia}, F. and {Danese}, L. and {Davies}, R.~D. and {Davis}, R.~J. and {de Bernardis}, P. and {de Rosa}, A. and {de Zotti}, G. and {Delabrouille}, J. and {D{\'e}sert}, F.-X. and {Dickinson}, C. and {Diego}, J.~M. and {Dole}, H. and {Donzelli}, S. and {Dor{\'e}}, O. and {Douspis}, M. and {Ducout}, A. and {Dupac}, X. and {Efstathiou}, G. and {Elsner}, F. and {En{\ss}lin}, T.~A. and {Eriksen}, H.~K. and {Falgarone}, E. and {Fergusson}, J. and {Finelli}, F. and {Forni}, O. and {Frailis}, M. and {Fraisse}, A.~A. and {Franceschi}, E. and {Frejsel}, A. and {Galeotta}, S. and {Galli}, S. and {Ganga}, K. and {Ghosh}, T. and {Giard}, M. and {Giraud-H{\'e}raud}, Y. and {Gjerl{\o}w}, E. and {Gonz{\'a}lez-Nuevo}, J. and {G{\'o}rski}, K.~M. and {Gratton}, S. and {Gregorio}, A. and {Gruppuso}, A. and {Gudmundsson}, J.~E. and {Hansen}, F.~K. and {Hanson}, D. and {Harrison}, D.~L. and {Helou}, G. and {Henrot-Versill{\'e}}, S. and {Hern{\'a}ndez-Monteagudo}, C. and {Herranz}, D. and {Hildebrandt}, S.~R. and {Hivon}, E. and {Hobson}, M. and {Holmes}, W.~A. and {Hornstrup}, A. and {Hovest}, W. and {Huffenberger}, K.~M. and {Hurier}, G. and {Jaffe}, A.~H. and {Jaffe}, T.~R. and {Jones}, W.~C. and {Juvela}, M. and {Keih{\"a}nen}, E. and {Keskitalo}, R. and {Kisner}, T.~S. and {Kneissl}, R. and {Knoche}, J. and {Kunz}, M. and {Kurki-Suonio}, H. and {Lagache}, G. and {L{\"a}hteenm{\"a}ki}, A. and {Lamarre}, J.-M. and {Lasenby}, A. and {Lattanzi}, M. and {Lawrence}, C.~R. and {Le Jeune}, M. and {Leahy}, J.~P. and {Leonardi}, R. and {Lesgourgues}, J. and {Levrier}, F. and {Liguori}, M. and {Lilje}, P.~B. and {Linden-V{\o}rnle}, M. and {L{\'o}pez-Caniego}, M. and {Lubin}, P.~M. and {Mac{\'\i}as-P{\'e}rez}, J.~F. and {Maggio}, G. and {Maino}, D. and {Mandolesi}, N. and {Mangilli}, A. and {Maris}, M. and {Marshall}, D.~J. and {Martin}, P.~G. and {Mart{\'\i}nez-Gonz{\'a}lez}, E. and {Masi}, S. and {Matarrese}, S. and {McGehee}, P. and {Meinhold}, P.~R. and {Melchiorri}, A. and {Mendes}, L. and {Mennella}, A. and {Migliaccio}, M. and {Mitra}, S. and {Miville-Desch{\^e}nes}, M.-A. and {Moneti}, A. and {Montier}, L. and {Morgante}, G. and {Mortlock}, D. and {Moss}, A. and {Munshi}, D. and {Murphy}, J.~A. and {Naselsky}, P. and {Nati}, F. and {Natoli}, P. and {Netterfield}, C.~B. and {N{\o}rgaard-Nielsen}, H.~U. and {Noviello}, F. and {Novikov}, D. and {Novikov}, I. and {Orlando}, E. and {Oxborrow}, C.~A. and {Paci}, F. and {Pagano}, L. and {Pajot}, F. and {Paladini}, R. and {Paoletti}, D. and {Partridge}, B. and {Pasian}, F. and {Patanchon}, G. and {Pearson}, T.~J. and {Perdereau}, O. and {Perotto}, L. and {Perrotta}, F. and {Pettorino}, V. and {Piacentini}, F. and {Piat}, M. and {Pierpaoli}, E. and {Pietrobon}, D. and {Plaszczynski}, S. and {Pointecouteau}, E. and {Polenta}, G. and {Pratt}, G.~W. and {Pr{\'e}zeau}, G. and {Prunet}, S. and {Puget}, J.-L. and {Rachen}, J.~P. and {Reach}, W.~T. and {Rebolo}, R. and {Reinecke}, M. and {Remazeilles}, M. and {Renault}, C. and {Renzi}, A. and {Ristorcelli}, I. and {Rocha}, G. and {Rosset}, C.},
        title = "{Planck 2015 results. X. Diffuse component separation: Foreground maps}",
      journal = {\aap},
         year = 2016,
        month = sep,
       volume = {594},
          eid = {A10},
        pages = {A10},
          doi = {10.1051/0004-6361/201525967},
archivePrefix = {arXiv},
       eprint = {1502.01588},
 primaryClass = {astro-ph.CO},
       adsurl = {https://ui.adsabs.harvard.edu/abs/2016A&A...594A..10P}
}

@ARTICLE{Planck_2020A&A...641A...2P,
       author = {{Planck Collaboration} and {Akrami}, Y. and {Arg{\"u}eso}, F. and {Ashdown}, M. and {Aumont}, J. and {Baccigalupi}, C. and {Ballardini}, M. and {Banday}, A.~J. and {Barreiro}, R.~B. and {Bartolo}, N. and {Basak}, S. and {Benabed}, K. and {Bernard}, J.-P. and {Bersanelli}, M. and {Bielewicz}, P. and {Bonavera}, L. and {Bond}, J.~R. and {Borrill}, J. and {Bouchet}, F.~R. and {Boulanger}, F. and {Bucher}, M. and {Burigana}, C. and {Butler}, R.~C. and {Calabrese}, E. and {Cardoso}, J.-F. and {Colombo}, L.~P.~L. and {Crill}, B.~P. and {Cuttaia}, F. and {de Bernardis}, P. and {de Rosa}, A. and {de Zotti}, G. and {Delabrouille}, J. and {Di Valentino}, E. and {Dickinson}, C. and {Diego}, J.~M. and {Donzelli}, S. and {Ducout}, A. and {Dupac}, X. and {Efstathiou}, G. and {Elsner}, F. and {En{\ss}lin}, T.~A. and {Eriksen}, H.~K. and {Fantaye}, Y. and {Finelli}, F. and {Frailis}, M. and {Franceschi}, E. and {Frolov}, A. and {Galeotta}, S. and {Galli}, S. and {Ganga}, K. and {G{\'e}nova-Santos}, R.~T. and {Gerbino}, M. and {Ghosh}, T. and {Gonz{\'a}lez-Nuevo}, J. and {G{\'o}rski}, K.~M. and {Gratton}, S. and {Gruppuso}, A. and {Gudmundsson}, J.~E. and {Handley}, W. and {Hansen}, F.~K. and {Herranz}, D. and {Hivon}, E. and {Huang}, Z. and {Jaffe}, A.~H. and {Jones}, W.~C. and {Karakci}, A. and {Keih{\"a}nen}, E. and {Keskitalo}, R. and {Kiiveri}, K. and {Kim}, J. and {Kisner}, T.~S. and {Krachmalnicoff}, N. and {Kunz}, M. and {Kurki-Suonio}, H. and {Lamarre}, J.-M. and {Lasenby}, A. and {Lattanzi}, M. and {Lawrence}, C.~R. and {Leahy}, J.~P. and {Levrier}, F. and {Liguori}, M. and {Lilje}, P.~B. and {Lindholm}, V. and {L{\'o}pez-Caniego}, M. and {Ma}, Y.-Z. and {Mac{\'\i}as-P{\'e}rez}, J.~F. and {Maggio}, G. and {Maino}, D. and {Mandolesi}, N. and {Mangilli}, A. and {Maris}, M. and {Martin}, P.~G. and {Mart{\'\i}nez-Gonz{\'a}lez}, E. and {Matarrese}, S. and {Mauri}, N. and {McEwen}, J.~D. and {Meinhold}, P.~R. and {Melchiorri}, A. and {Mennella}, A. and {Migliaccio}, M. and {Molinari}, D. and {Montier}, L. and {Morgante}, G. and {Moss}, A. and {Natoli}, P. and {Pagano}, L. and {Paoletti}, D. and {Partridge}, B. and {Patanchon}, G. and {Patrizii}, L. and {Peel}, M. and {Perrotta}, F. and {Pettorino}, V. and {Piacentini}, F. and {Polenta}, G. and {Puget}, J.-L. and {Rachen}, J.~P. and {Racine}, B. and {Reinecke}, M. and {Remazeilles}, M. and {Renzi}, A. and {Rocha}, G. and {Roudier}, G. and {Rubi{\~n}o-Mart{\'\i}n}, J.~A. and {Salvati}, L. and {Sandri}, M. and {Savelainen}, M. and {Scott}, D. and {Seljebotn}, D.~S. and {Sirignano}, C. and {Sirri}, G. and {Spencer}, L.~D. and {Suur-Uski}, A.-S. and {Tauber}, J.~A. and {Tavagnacco}, D. and {Tenti}, M. and {Terenzi}, L. and {Toffolatti}, L. and {Tomasi}, M. and {Trombetti}, T. and {Valiviita}, J. and {Vansyngel}, F. and {Van Tent}, B. and {Vielva}, P. and {Villa}, F. and {Vittorio}, N. and {Wandelt}, B.~D. and {Watson}, R. and {Wehus}, I.~K. and {Zacchei}, A. and {Zonca}, A.},
        title = "{Planck 2018 results. II. Low Frequency Instrument data processing}",
      journal = {\aap},
         year = 2020,
        month = sep,
       volume = {641},
          eid = {A2},
        pages = {A2},
          doi = {10.1051/0004-6361/201833293},
archivePrefix = {arXiv},
       eprint = {1807.06206},
 primaryClass = {astro-ph.CO},
       adsurl = {https://ui.adsabs.harvard.edu/abs/2020A&A...641A...2P}
}

@ARTICLE{Popescu_2017,
       author = {{Popescu}, C.~C. and {Yang}, R. and {Tuffs}, R.~J. and {Natale}, G. and {Rushton}, M. and {Aharonian}, F.},
        title = "{A radiation transfer model for the Milky Way: I. Radiation fields and application to high-energy astrophysics★}",
      journal = {\mnras},
         year = 2017,
        month = sep,
       volume = {470},
       number = {3},
        pages = {2539-2558},
          doi = {10.1093/mnras/stx1282},
archivePrefix = {arXiv},
       eprint = {1705.06652},
 primaryClass = {astro-ph.GA},
       adsurl = {https://ui.adsabs.harvard.edu/abs/2017MNRAS.470.2539P}
}

@ARTICLE{Clark_2020,
       author = {{Clark}, J.~S. and {Ritchie}, B.~W. and {Negueruela}, I.},
        title = "{A VLT/FLAMES survey for massive binaries in Westerlund 1. VII. Cluster census}",
      journal = {\aap},
         year = 2020,
        month = mar,
       volume = {635},
          eid = {A187},
        pages = {A187},
          doi = {10.1051/0004-6361/201935903},
archivePrefix = {arXiv},
       eprint = {1908.05616},
 primaryClass = {astro-ph.SR},
       adsurl = {https://ui.adsabs.harvard.edu/abs/2020A&A...635A.187C}
}

@ARTICLE{Crowther_2006MNRAS.372.1407C,
       author = {{Crowther}, Paul A. and {Hadfield}, L.~J. and {Clark}, J.~S. and {Negueruela}, I. and {Vacca}, W.~D.},
        title = "{A census of the Wolf-Rayet content in Westerlund 1 from near-infrared imaging and spectroscopy}",
      journal = {\mnras},
         year = 2006,
        month = nov,
       volume = {372},
       number = {3},
        pages = {1407-1424},
          doi = {10.1111/j.1365-2966.2006.10952.x},
archivePrefix = {arXiv},
       eprint = {astro-ph/0608356},
 primaryClass = {astro-ph},
       adsurl = {https://ui.adsabs.harvard.edu/abs/2006MNRAS.372.1407C}
}

@ARTICLE{Martins_2005A&A...436.1049M,
       author = {{Martins}, F. and {Schaerer}, D. and {Hillier}, D.~J.},
        title = "{A new calibration of stellar parameters of Galactic O stars}",
      journal = {\aap},
         year = 2005,
        month = jun,
       volume = {436},
       number = {3},
        pages = {1049-1065},
          doi = {10.1051/0004-6361:20042386},
archivePrefix = {arXiv},
       eprint = {astro-ph/0503346},
 primaryClass = {astro-ph},
       adsurl = {https://ui.adsabs.harvard.edu/abs/2005A&A...436.1049M}
}

@ARTICLE{Crowther_2006A&A...446..279C,
       author = {{Crowther}, P.~A. and {Lennon}, D.~J. and {Walborn}, N.~R.},
        title = "{Physical parameters and wind properties of galactic early B supergiants}",
      journal = {\aap},
         year = 2006,
        month = jan,
       volume = {446},
       number = {1},
        pages = {279-293},
          doi = {10.1051/0004-6361:20053685},
archivePrefix = {arXiv},
       eprint = {astro-ph/0509436},
 primaryClass = {astro-ph},
       adsurl = {https://ui.adsabs.harvard.edu/abs/2006A&A...446..279C}
}

@ARTICLE{erosita_2025,
       author = {{Haubner}, Konstantin and {Sasaki}, Manami and {Mitchell}, Alison and {Ponti}, Gabriele and {Rowell}, Gavin and {Einecke}, Sabrina and {Filipovi{\'c}}, Miroslav and {Lazarevi{\'c}}, Sanja and {P{\"u}hlhofer}, Gerd and {Strong}, Andrew},
        title = "{eROSITA X-ray analysis of the PeVatron candidate Westerlund 1}",
      journal = {\aap},
         year = 2025,
        month = mar,
       volume = {695},
          eid = {A3},
        pages = {A3},
          doi = {10.1051/0004-6361/202451964},
archivePrefix = {arXiv},
       eprint = {2501.12990},
 primaryClass = {astro-ph.HE},
       adsurl = {https://ui.adsabs.harvard.edu/abs/2025A&A...695A...3H}
}

@ARTICLE{HESS_2018A&A...612A...1H,
       author = {{H.~E.~S.~S. Collaboration} and {Abdalla}, H. and {Abramowski}, A. and {Aharonian}, F. and {Ait Benkhali}, F. and {Ang{\"u}ner}, E.~O. and {Arakawa}, M. and {Arrieta}, M. and {Aubert}, P. and {Backes}, M. and {Balzer}, A. and {Barnard}, M. and {Becherini}, Y. and {Becker Tjus}, J. and {Berge}, D. and {Bernhard}, S. and {Bernl{\"o}hr}, K. and {Blackwell}, R. and {B{\"o}ttcher}, M. and {Boisson}, C. and {Bolmont}, J. and {Bonnefoy}, S. and {Bordas}, P. and {Bregeon}, J. and {Brun}, F. and {Brun}, P. and {Bryan}, M. and {B{\"u}chele}, M. and {Bulik}, T. and {Capasso}, M. and {Carrigan}, S. and {Caroff}, S. and {Carosi}, A. and {Casanova}, S. and {Cerruti}, M. and {Chakraborty}, N. and {Chaves}, R.~C.~G. and {Chen}, A. and {Chevalier}, J. and {Colafrancesco}, S. and {Condon}, B. and {Conrad}, J. and {Davids}, I.~D. and {Decock}, J. and {Deil}, C. and {Devin}, J. and {deWilt}, P. and {Dirson}, L. and {Djannati-Ata{\"\i}}, A. and {Domainko}, W. and {Donath}, A. and {Drury}, L.~O. 'C. and {Dutson}, K. and {Dyks}, J. and {Edwards}, T. and {Egberts}, K. and {Eger}, P. and {Emery}, G. and {Ernenwein}, J.-P. and {Eschbach}, S. and {Farnier}, C. and {Fegan}, S. and {Fernandes}, M.~V. and {Fiasson}, A. and {Fontaine}, G. and {F{\"o}rster}, A. and {Funk}, S. and {F{\"u}{\ss}ling}, M. and {Gabici}, S. and {Gallant}, Y.~A. and {Garrigoux}, T. and {Gast}, H. and {Gat{\'e}}, F. and {Giavitto}, G. and {Giebels}, B. and {Glawion}, D. and {Glicenstein}, J.~F. and {Gottschall}, D. and {Grondin}, M.-H. and {Hahn}, J. and {Haupt}, M. and {Hawkes}, J. and {Heinzelmann}, G. and {Henri}, G. and {Hermann}, G. and {Hinton}, J.~A. and {Hofmann}, W. and {Hoischen}, C. and {Holch}, T.~L. and {Holler}, M. and {Horns}, D. and {Ivascenko}, A. and {Iwasaki}, H. and {Jacholkowska}, A. and {Jamrozy}, M. and {Jankowsky}, D. and {Jankowsky}, F. and {Jingo}, M. and {Jouvin}, L. and {Jung-Richardt}, I. and {Kastendieck}, M.~A. and {Katarzy{\'n}ski}, K. and {Katsuragawa}, M. and {Katz}, U. and {Kerszberg}, D. and {Khangulyan}, D. and {Kh{\'e}lifi}, B. and {King}, J. and {Klepser}, S. and {Klochkov}, D. and {Klu{\'z}niak}, W. and {Komin}, Nu. and {Kosack}, K. and {Krakau}, S. and {Kraus}, M. and {Kr{\"u}ger}, P.~P. and {Laffon}, H. and {Lamanna}, G. and {Lau}, J. and {Lees}, J.-P. and {Lefaucheur}, J. and {Lemi{\`e}re}, A. and {Lemoine-Goumard}, M. and {Lenain}, J.-P. and {Leser}, E. and {Lohse}, T. and {Lorentz}, M. and {Liu}, R. and {L{\'o}pez-Coto}, R. and {Lypova}, I. and {Marandon}, V. and {Malyshev}, D. and {Marcowith}, A. and {Mariaud}, C. and {Marx}, R. and {Maurin}, G. and {Maxted}, N. and {Mayer}, M. and {Meintjes}, P.~J. and {Meyer}, M. and {Mitchell}, A.~M.~W. and {Moderski}, R. and {Mohamed}, M. and {Mohrmann}, L. and {Mor{\r{a}}}, K. and {Moulin}, E. and {Murach}, T. and {Nakashima}, S. and {de Naurois}, M. and {Ndiyavala}, H. and {Niederwanger}, F. and {Niemiec}, J. and {Oakes}, L. and {O'Brien}, P. and {Odaka}, H. and {Ohm}, S. and {Ostrowski}, M. and {Oya}, I. and {Padovani}, M. and {Panter}, M. and {Parsons}, R.~D. and {Paz Arribas}, M. and {Pekeur}, N.~W. and {Pelletier}, G. and {Perennes}, C. and {Petrucci}, P.-O. and {Peyaud}, B. and {Piel}, Q. and {Pita}, S. and {Poireau}, V. and {Poon}, H. and {Prokhorov}, D. and {Prokoph}, H. and {P{\"u}hlhofer}, G. and {Punch}, M. and {Quirrenbach}, A. and {Raab}, S. and {Rauth}, R. and {Reimer}, A. and {Reimer}, O. and {Renaud}, M. and {de los Reyes}, R. and {Rieger}, F. and {Rinchiuso}, L. and {Romoli}, C. and {Rowell}, G. and {Rudak}, B. and {Rulten}, C.~B. and {Safi-Harb}, S. and {Sahakian}, V. and {Saito}, S. and {Sanchez}, D.~A. and {Santangelo}, A. and {Sasaki}, M. and {Schandri}, M. and {Schlickeiser}, R. and {Sch{\"u}ssler}, F. and {Schulz}, A. and {Schwanke}, U. and {Schwemmer}, S.},
        title = "{The H.E.S.S. Galactic plane survey}",
      journal = {\aap},
         year = 2018,
        month = apr,
       volume = {612},
          eid = {A1},
        pages = {A1},
          doi = {10.1051/0004-6361/201732098},
archivePrefix = {arXiv},
       eprint = {1804.02432},
 primaryClass = {astro-ph.HE},
       adsurl = {https://ui.adsabs.harvard.edu/abs/2018A&A...612A...1H}
}

@ARTICLE{IceCube_2015EPJC...75..116A,
       author = {{Aartsen}, M.~G. and {Ackermann}, M. and {Adams}, J. and {Aguilar}, J.~A. and {Ahlers}, M. and {Ahrens}, M. and {Altmann}, D. and {Anderson}, T. and {Arguelles}, C. and {Arlen}, T.~C. and {Auffenberg}, J. and {Bai}, X. and {Barwick}, S.~W. and {Baum}, V. and {Beatty}, J.~J. and {Tjus}, J. Becker and {Becker}, K.-H. and {BenZvi}, S. and {Berghaus}, P. and {Berley}, D. and {Bernardini}, E. and {Bernhard}, A. and {Besson}, D.~Z. and {Binder}, G. and {Bindig}, D. and {Bissok}, M. and {Blaufuss}, E. and {Blumenthal}, J. and {Boersma}, D.~J. and {Bohm}, C. and {Bos}, F. and {Bose}, D. and {B{\"o}ser}, S. and {Botner}, O. and {Brayeur}, L. and {Bretz}, H.~P. and {Brown}, A.~M. and {Casey}, J. and {Casier}, M. and {Cheung}, E. and {Chirkin}, D. and {Christov}, A. and {Christy}, B. and {Clark}, K. and {Classen}, L. and {Clevermann}, F. and {Coenders}, S. and {Cowen}, D.~F. and {Cruz Silva}, A.~H. and {Danninger}, M. and {Daughhetee}, J. and {Davis}, J.~C. and {Day}, M. and {de Andr{\'e}}, J.~P.~A.~M. and {De Clercq}, C. and {De Ridder}, S. and {Desiati}, P. and {de Vries}, K.~D. and {de With}, M. and {DeYoung}, T. and {D{\'\i}az-V{\'e}lez}, J.~C. and {Dunkman}, M. and {Eagan}, R. and {Eberhardt}, B. and {Eichmann}, B. and {Eisch}, J. and {Euler}, S. and {Evenson}, P.~A. and {Fadiran}, O. and {Fazely}, A.~R. and {Fedynitch}, A. and {Feintzeig}, J. and {Felde}, J. and {Feusels}, T. and {Filimonov}, K. and {Finley}, C. and {Fischer-Wasels}, T. and {Flis}, S. and {Franckowiak}, A. and {Frantzen}, K. and {Fuchs}, T. and {Gaisser}, T.~K. and {Gaior}, R. and {Gallagher}, J. and {Gerhardt}, L. and {Gier}, D. and {Gladstone}, L. and {Gl{\"u}senkamp}, T. and {Goldschmidt}, A. and {Golup}, G. and {Gonzalez}, J.~G. and {Goodman}, J.~A. and {G{\'o}ra}, D. and {Grant}, D. and {Gretskov}, P. and {Groh}, J.~C. and {Gro{\ss}}, A. and {Ha}, C. and {Haack}, C. and {Haj Ismail}, A. and {Hallen}, P. and {Hallgren}, A. and {Halzen}, F. and {Hanson}, K. and {Hebecker}, D. and {Heereman}, D. and {Heinen}, D. and {Helbing}, K. and {Hellauer}, R. and {Hellwig}, D. and {Hickford}, S. and {Hill}, G.~C. and {Hoffman}, K.~D. and {Hoffmann}, R. and {Homeier}, A. and {Hoshina}, K. and {Huang}, F. and {Huelsnitz}, W. and {Hulth}, P.~O. and {Hultqvist}, K. and {Hussain}, S. and {Ishihara}, A. and {Jacobi}, E. and {Jacobsen}, J. and {Jagielski}, K. and {Japaridze}, G.~S. and {Jero}, K. and {Jlelati}, O. and {Jurkovic}, M. and {Kaminsky}, B. and {Kappes}, A. and {Karg}, T. and {Karle}, A. and {Kauer}, M. and {Keivani}, A. and {Kelley}, J.~L. and {Kheirandish}, A. and {Kiryluk}, J. and {Kl{\"a}s}, J. and {Klein}, S.~R. and {K{\"o}hne}, J.~H. and {Kohnen}, G. and {Kolanoski}, H. and {Koob}, A. and {K{\"o}pke}, L. and {Kopper}, C. and {Kopper}, S. and {Koskinen}, D.~J. and {Kowalski}, M. and {Kriesten}, A. and {Krings}, K. and {Kroll}, G. and {Kroll}, M. and {Kunnen}, J. and {Kurahashi}, N. and {Kuwabara}, T. and {Labare}, M. and {Larsen}, D.~T. and {Larson}, M.~J. and {Lesiak-Bzdak}, M. and {Leuermann}, M. and {Leute}, J. and {L{\"u}nemann}, J. and {Madsen}, J. and {Maggi}, G. and {Maruyama}, R. and {Mase}, K. and {Matis}, H.~S. and {Maunu}, R. and {McNally}, F. and {Meagher}, K. and {Medici}, M. and {Meli}, A. and {Meures}, T. and {Miarecki}, S. and {Middell}, E. and {Middlemas}, E. and {Milke}, N. and {Miller}, J. and {Mohrmann}, L. and {Montaruli}, T. and {Morse}, R. and {Nahnhauer}, R. and {Naumann}, U. and {Niederhausen}, H. and {Nowicki}, S.~C. and {Nygren}, D.~R. and {Obertacke}, A. and {Odrowski}, S. and {Olivas}, A. and {Omairat}, A. and {O'Murchadha}, A. and {Palczewski}, T. and {Paul}, L. and {Penek}, {\"O}. and {Pepper}, J.~A. and {P{\'e}rez de los Heros}, C. and {Pfendner}, C. and {Pieloth}, D. and {Pinat}, E.},
        title = "{Development of a general analysis and unfolding scheme and its application to measure the energy spectrum of atmospheric neutrinos with IceCube: IceCube Collaboration}",
      journal = {European Physical Journal C},
         year = 2015,
        month = mar,
       volume = {75},
          eid = {116},
        pages = {116},
          doi = {10.1140/epjc/s10052-015-3330-z},
archivePrefix = {arXiv},
       eprint = {1409.4535},
 primaryClass = {astro-ph.HE},
       adsurl = {https://ui.adsabs.harvard.edu/abs/2015EPJC...75..116A}
}

@ARTICLE{Mascaretti_2020APh...114...22M,
       author = {{Mascaretti}, Carlo and {Blasi}, Pasquale and {Evoli}, Carmelo},
        title = "{Atmospheric neutrinos and the knee of the cosmic ray spectrum}",
      journal = {Astroparticle Physics},
         year = 2020,
        month = jan,
       volume = {114},
        pages = {22-29},
          doi = {10.1016/j.astropartphys.2019.06.002},
archivePrefix = {arXiv},
       eprint = {1906.05197},
 primaryClass = {astro-ph.HE},
       adsurl = {https://ui.adsabs.harvard.edu/abs/2020APh...114...22M}
}

@ARTICLE{km3net_2016JPhG...43h4001A,
       author = {{Adri{\'a}n-Mart{\'\i}nez}, S. and {Ageron}, M. and {Aharonian}, F. and {Aiello}, S. and {Albert}, A. and {Ameli}, F. and {Anassontzis}, E. and {Andre}, M. and {Androulakis}, G. and {Anghinolfi}, M. and {Anton}, G. and {Ardid}, M. and {Avgitas}, T. and {Barbarino}, G. and {Barbarito}, E. and {Baret}, B. and {Barrios-Mart{\'\i}}, J. and {Belhorma}, B. and {Belias}, A. and {Berbee}, E. and {van den Berg}, A. and {Bertin}, V. and {Beurthey}, S. and {van Beveren}, V. and {Beverini}, N. and {Biagi}, S. and {Biagioni}, A. and {Billault}, M. and {Bond{\`\i}}, M. and {Bormuth}, R. and {Bouhadef}, B. and {Bourlis}, G. and {Bourret}, S. and {Boutonnet}, C. and {Bouwhuis}, M. and {Bozza}, C. and {Bruijn}, R. and {Brunner}, J. and {Buis}, E. and {Busto}, J. and {Cacopardo}, G. and {Caillat}, L. and {Calamai}, M. and {Calvo}, D. and {Capone}, A. and {Caramete}, L. and {Cecchini}, S. and {Celli}, S. and {Champion}, C. and {Cherkaoui El Moursli}, R. and {Cherubini}, S. and {Chiarusi}, T. and {Circella}, M. and {Classen}, L. and {Cocimano}, R. and {Coelho}, J.~A.~B. and {Coleiro}, A. and {Colonges}, S. and {Coniglione}, R. and {Cordelli}, M. and {Cosquer}, A. and {Coyle}, P. and {Creusot}, A. and {Cuttone}, G. and {D'Amico}, A. and {De Bonis}, G. and {De Rosa}, G. and {De Sio}, C. and {Di Capua}, F. and {Di Palma}, I. and {D{\'\i}az Garc{\'\i}a}, A.~F. and {Distefano}, C. and {Donzaud}, C. and {Dornic}, D. and {Dorosti-Hasankiadeh}, Q. and {Drakopoulou}, E. and {Drouhin}, D. and {Drury}, L. and {Durocher}, M. and {Eberl}, T. and {Eichie}, S. and {van Eijk}, D. and {El Bojaddaini}, I. and {El Khayati}, N. and {Elsaesser}, D. and {Enzenh{\"o}fer}, A. and {Fassi}, F. and {Favali}, P. and {Fermani}, P. and {Ferrara}, G. and {Filippidis}, C. and {Frascadore}, G. and {Fusco}, L.~A. and {Gal}, T. and {Galat{\`a}}, S. and {Garufi}, F. and {Gay}, P. and {Gebyehu}, M. and {Giordano}, V. and {Gizani}, N. and {Gracia}, R. and {Graf}, K. and {Gr{\'e}goire}, T. and {Grella}, G. and {Habel}, R. and {Hallmann}, S. and {van Haren}, H. and {Harissopulos}, S. and {Heid}, T. and {Heijboer}, A. and {Heine}, E. and {Henry}, S. and {Hern{\'a}ndez-Rey}, J.~J. and {Hevinga}, M. and {Hofest{\"a}dt}, J. and {Hugon}, C.~M.~F. and {Illuminati}, G. and {James}, C.~W. and {Jansweijer}, P. and {Jongen}, M. and {de Jong}, M. and {Kadler}, M. and {Kalekin}, O. and {Kappes}, A. and {Katz}, U.~F. and {Keller}, P. and {Kieft}, G. and {Kie{\ss}ling}, D. and {Koffeman}, E.~N. and {Kooijman}, P. and {Kouchner}, A. and {Kulikovskiy}, V. and {Lahmann}, R. and {Lamare}, P. and {Leisos}, A. and {Leonora}, E. and {Clark}, M. Lindsey and {Liolios}, A. and {Llorens Alvarez}, C.~D. and {Lo Presti}, D. and {L{\"o}hner}, H. and {Lonardo}, A. and {Lotze}, M. and {Loucatos}, S. and {Maccioni}, E. and {Mannheim}, K. and {Margiotta}, A. and {Marinelli}, A. and {Mari{\textcommabelow s}}, O. and {Markou}, C. and {Mart{\'\i}nez-Mora}, J.~A. and {Martini}, A. and {Mele}, R. and {Melis}, K.~W. and {Michael}, T. and {Migliozzi}, P. and {Migneco}, E. and {Mijakowski}, P. and {Miraglia}, A. and {Mollo}, C.~M. and {Mongelli}, M. and {Morganti}, M. and {Moussa}, A. and {Musico}, P. and {Musumeci}, M. and {Navas}, S. and {Nicolau}, C.~A. and {Olcina}, I. and {Olivetto}, C. and {Orlando}, A. and {Papaikonomou}, A. and {Papaleo}, R. and {P{\u{a}}v{\u{a}}la{\textcommabelow s}}, G.~E. and {Peek}, H. and {Pellegrino}, C. and {Perrina}, C. and {Pfutzner}, M. and {Piattelli}, P. and {Pikounis}, K. and {Poma}, G.~E. and {Popa}, V. and {Pradier}, T. and {Pratolongo}, F. and {P{\"u}hlhofer}, G. and {Pulvirenti}, S. and {Quinn}, L. and {Racca}, C. and {Raffaelli}, F. and {Randazzo}, N. and {Rapidis}, P. and {Razis}, P. and {Real}, D. and {Resvanis}, L. and {Reubelt}, J. and {Riccobene}, G. and {Rossi}, C. and {Rovelli}, A. and {Salda{\~n}a}, M. and {Salvadori}, I. and {Samtleben}, D.~F.~E.},
        title = "{Letter of intent for KM3NeT 2.0}",
      journal = {Journal of Physics G Nuclear Physics},
         year = 2016,
        month = aug,
       volume = {43},
       number = {8},
          eid = {084001},
        pages = {084001},
          doi = {10.1088/0954-3899/43/8/084001},
archivePrefix = {arXiv},
       eprint = {1601.07459},
 primaryClass = {astro-ph.IM},
       adsurl = {https://ui.adsabs.harvard.edu/abs/2016JPhG...43h4001A}
}

@ARTICLE{km3net_2024EPJC...84..885K,
       author = {{KM3NeT Collaboration} and {Aiello}, S. and {Albert}, A. and {Alshamsi}, M. and {Alves Garre}, S. and {Aly}, Z. and {Ambrosone}, A. and {Ameli}, F. and {Andre}, M. and {Androutsou}, E. and {Anguita}, M. and {Aphecetche}, L. and {Ardid}, M. and {Ardid}, S. and {Atmani}, H. and {Aublin}, J. and {Badaracco}, F. and {Bailly-Salins}, L. and {Barda{\v{c}}ov{\'a}}, Z. and {Baret}, B. and {Bariego-Quintana}, A. and {Baruzzi}, A. and {Basegmez du Pree}, S. and {Becherini}, Y. and {Bendahman}, M. and {Benfenati}, F. and {Benhassi}, M. and {Benoit}, D.~M. and {Berbee}, E. and {Bertin}, V. and {Biagi}, S. and {Boettcher}, M. and {Bonanno}, D. and {Boumaaza}, J. and {Bouta}, M. and {Bouwhuis}, M. and {Bozza}, C. and {Bozza}, R.~M. and {Br{\^a}nza{\textcommabelow s}}, H. and {Bretaudeau}, F. and {Breuhaus}, M. and {Bruijn}, R. and {Brunner}, J. and {Bruno}, R. and {Buis}, E. and {Buompane}, R. and {Busto}, J. and {Caiffi}, B. and {Calvo}, D. and {Campion}, S. and {Capone}, A. and {Carenini}, F. and {Carretero}, V. and {Cartraud}, T. and {Castaldi}, P. and {Cecchini}, V. and {Celli}, S. and {Cerisy}, L. and {Chabab}, M. and {Chadolias}, M. and {Chen}, A. and {Cherubini}, S. and {Chiarusi}, T. and {Circella}, M. and {Cocimano}, R. and {Coelho}, J.~A.~B. and {Coleiro}, A. and {Coniglione}, R. and {Coyle}, P. and {Creusot}, A. and {Cuttone}, G. and {Dallier}, R. and {Darras}, Y. and {De Benedittis}, A. and {De Martino}, B. and {Decoene}, V. and {Del Burgo}, R. and {Del Rosso}, I. and {Di Mauro}, L.~S. and {Di Palma}, I. and {D{\'\i}az}, A.~F. and {Diaz}, C. and {Diego-Tortosa}, D. and {Distefano}, C. and {Domi}, A. and {Donzaud}, C. and {Dornic}, D. and {D{\"o}rr}, M. and {Drakopoulou}, E. and {Drouhin}, D. and {Ducoin}, J.-G. and {Dvornick{\'y}}, R. and {Eberl}, T. and {Eckerov{\'a}}, E. and {Eddymaoui}, A. and {van Eeden}, T. and {Eff}, M. and {van Eijk}, D. and {El Bojaddaini}, I. and {El Hedri}, S. and {Enzenh{\"o}fer}, A. and {Ferrara}, G. and {Filipovi{\'c}}, M.~D. and {Filippini}, F. and {Franciotti}, D. and {Fusco}, L.~A. and {Gabriel}, J. and {Gagliardini}, S. and {Gal}, T. and {Garc{\'\i}a M{\'e}ndez}, J. and {Garcia Soto}, A. and {Gatius Oliver}, C. and {Gei{\ss}elbrecht}, N. and {Ghaddari}, H. and {Gialanella}, L. and {Gibson}, B.~K. and {Giorgio}, E. and {Goos}, I. and {Goswami}, P. and {Goupilliere}, D. and {Gozzini}, S.~R. and {Gracia}, R. and {Graf}, K. and {Guidi}, C. and {Guillon}, B. and {Guti{\'e}rrez}, M. and {van Haren}, H. and {Heijboer}, A. and {Hekalo}, A. and {Hennig}, L. and {Hern{\'a}ndez-Rey}, J.~J. and {Idrissi Ibnsalih}, W. and {Illuminati}, G. and {de Jong}, M. and {de Jong}, P. and {Jung}, B.~J. and {Kalaczy{\'n}ski}, P. and {Kalekin}, O. and {Katz}, U.~F. and {Kistauri}, G. and {Kopper}, C. and {Kouchner}, A. and {Kueviakoe}, V. and {Kulikovskiy}, V. and {Kvatadze}, R. and {Labalme}, M. and {Lahmann}, R. and {Larosa}, G. and {Lastoria}, C. and {Lazo}, A. and {Le Stum}, S. and {Lehaut}, G. and {Leonora}, E. and {Lessing}, N. and {Levi}, G. and {Lindsey Clark}, M. and {Longhitano}, F. and {Magnani}, F. and {Majumdar}, J. and {Malerba}, L. and {Mamedov}, F. and {Ma{\'n}czak}, J. and {Manfreda}, A. and {Marconi}, M. and {Margiotta}, A. and {Marinelli}, A. and {Markou}, C. and {Martin}, L. and {Mart{\'\i}nez-Mora}, J.~A. and {Marzaioli}, F. and {Mastrodicasa}, M. and {Mastroianni}, S. and {Miccich{\`e}}, S. and {Miele}, G. and {Migliozzi}, P. and {Migneco}, E. and {Mitsou}, M.~L. and {Mollo}, C.~M. and {Morales-Gallegos}, L. and {Morga}, M. and {Moussa}, A. and {Mozun Mateo}, I. and {Muller}, R. and {Musone}, M.~R. and {Musumeci}, M. and {Navas}, S. and {Nayerhoda}, A. and {Nicolau}, C.~A. and {Nkosi}, B. and {{\'O} Fearraigh}, B. and {Oliviero}, V. and {Orlando}, A. and {Oukacha}, E. and {Paesani}, D. and {Palacios Gonz{\'a}lez}, J. and {Papalashvili}, G. and {Parisi}, V. and {Pastor Gomez}, E.~J. and {P{\u{a}}un}, A.~M. and {P{\u{a}}v{\u{a}}la{\textcommabelow s}}, G.~E.},
        title = "{Astronomy potential of KM3NeT/ARCA}",
      journal = {European Physical Journal C},
         year = 2024,
        month = sep,
       volume = {84},
       number = {9},
          eid = {885},
        pages = {885},
          doi = {10.1140/epjc/s10052-024-13137-2},
archivePrefix = {arXiv},
       eprint = {2402.08363},
 primaryClass = {astro-ph.HE},
       adsurl = {https://ui.adsabs.harvard.edu/abs/2024EPJC...84..885K}
}

@ARTICLE{km3net_2019APh...111..100A,
       author = {{Aiello}, S. and {Akrame}, S.~E. and {Ameli}, F. and {Anassontzis}, E.~G. and {Andre}, M. and {Androulakis}, G. and {Anghinolfi}, M. and {Anton}, G. and {Ardid}, M. and {Aublin}, J. and {Avgitas}, T. and {Bagatelas}, C. and {Barbarino}, G. and {Baret}, B. and {Barrios-Mart{\'\i}}, J. and {Belias}, A. and {Berbee}, E. and {van den Berg}, A. and {Bertin}, V. and {Biagi}, S. and {Biagioni}, A. and {Biernoth}, C. and {Boumaaza}, J. and {Bourret}, S. and {Bouta}, M. and {Bouwhuis}, M. and {Bozza}, C. and {Br{\^a}nza{\textcommabelow s}}, H. and {Bruchner}, M. and {Bruijn}, R. and {Brunner}, J. and {Buis}, E. and {Buompane}, R. and {Busto}, J. and {Calvo}, D. and {Capone}, A. and {Celli}, S. and {Chabab}, M. and {Chau}, N. and {Cherubini}, S. and {Chiarella}, V. and {Chiarusi}, T. and {Circella}, M. and {Cocimano}, R. and {Coelho}, J.~A.~B. and {Coleiro}, A. and {Molla}, M. Colomer and {Coniglione}, R. and {Coyle}, P. and {Creusot}, A. and {Cuttone}, G. and {D'Onofrio}, A. and {Dallier}, R. and {De Sio}, C. and {Di Palma}, I. and {D{\'\i}az}, A.~F. and {Diego-Tortosa}, D. and {Distefano}, C. and {Domi}, A. and {Don{\`a}}, R. and {Donzaud}, C. and {Dornic}, D. and {D{\"o}rr}, M. and {Durocher}, M. and {Eberl}, T. and {van Eijk}, D. and {El Bojaddaini}, I. and {Eljarrari}, H. and {Elsaesser}, D. and {Enzenh{\"o}fer}, A. and {Fermani}, P. and {Ferrara}, G. and {Filipovi{\'c}}, M.~D. and {Fusco}, L.~A. and {Gal}, T. and {Garcia}, A. and {Garufi}, F. and {Gialanella}, L. and {Giorgio}, E. and {Giuliante}, A. and {Gozzini}, S.~R. and {Gracia}, R. and {Graf}, K. and {Grasso}, D. and {Gr{\'e}goire}, T. and {Grella}, G. and {Hallmann}, S. and {Hamdaoui}, H. and {van Haren}, H. and {Heid}, T. and {Heijboer}, A. and {Hekalo}, A. and {Hern{\'a}ndez-Rey}, J.~J. and {Hofest{\"a}dt}, J. and {Illuminati}, G. and {James}, C.~W. and {Jongen}, M. and {de Jong}, M. and {de Jong}, P. and {Kadler}, M. and {Kalaczy{\'n}ski}, P. and {Kalekin}, O. and {Katz}, U.~F. and {Khan Chowdhury}, N.~R. and {Kie{\ss}ling}, D. and {Koffeman}, E.~N. and {Kooijman}, P. and {Kouchner}, A. and {Kreter}, M. and {Kulikovskiy}, V. and {Kunhikannan Kannichankandy}, M. and {Lahmann}, R. and {Larosa}, G. and {Le Breton}, R. and {Leone}, F. and {Leonora}, E. and {Levi}, G. and {Lincetto}, M. and {Lonardo}, A. and {Longhitano}, F. and {Lopez Coto}, D. and {Lotze}, M. and {Maderer}, L. and {Maggi}, G. and {Ma{\'n}czak}, J. and {Mannheim}, K. and {Margiotta}, A. and {Marinelli}, A. and {Markou}, C. and {Martin}, L. and {Mart{\'\i}nez-Mora}, J.~A. and {Martini}, A. and {Marzaioli}, F. and {Mele}, R. and {Melis}, K.~W. and {Migliozzi}, P. and {Migneco}, E. and {Mijakowski}, P. and {Miranda}, L.~S. and {Mollo}, C.~M. and {Morganti}, M. and {Moser}, M. and {Moussa}, A. and {Muller}, R. and {Musumeci}, M. and {Nauta}, L. and {Navas}, S. and {Nicolau}, C.~A. and {Nielsen}, C. and {{\'O} Fearraigh}, B. and {Organokov}, M. and {Orlando}, A. and {Ottonello}, S. and {Panagopoulos}, V. and {Papalashvili}, G. and {Papaleo}, R. and {P{\u{a}}v{\u{a}}la{\textcommabelow s}}, G.~E. and {Pellegrino}, C. and {Perrin-Terrin}, M. and {Piattelli}, P. and {Pikounis}, K. and {Pisanti}, O. and {Poir{\'e}}, C. and {Polydefki}, G. and {Popa}, V. and {Post}, M. and {Pradier}, T. and {P{\"u}hlhofer}, G. and {Pulvirenti}, S. and {Quinn}, L. and {Raffaelli}, F. and {Randazzo}, N. and {Razzaque}, S. and {Real}, D. and {Resvanis}, L. and {Reubelt}, J. and {Riccobene}, G. and {Richer}, M. and {Rigalleau}, L. and {Rovelli}, A. and {Saffer}, M. and {Salvadori}, I. and {Samtleben}, D.~F.~E. and {S{\'a}nchez Losa}, A. and {Sanguineti}, M. and {Santangelo}, A. and {Santonocito}, D. and {Sapienza}, P. and {Schumann}, J. and {Sciacca}, V. and {Seneca}, J. and {Sgura}, I. and {Shanidze}, R. and {Sharma}, A. and {Simeone}, F. and {Sinopoulou}, A. and {Spisso}, B. and {Spurio}, M. and {Stavropoulos}, D. and {Steijger}, J.},
        title = "{Sensitivity of the KM3NeT/ARCA neutrino telescope to point-like neutrino sources}",
      journal = {Astroparticle Physics},
         year = 2019,
        month = sep,
       volume = {111},
        pages = {100-110},
          doi = {10.1016/j.astropartphys.2019.04.002},
archivePrefix = {arXiv},
       eprint = {1810.08499},
 primaryClass = {astro-ph.HE},
       adsurl = {https://ui.adsabs.harvard.edu/abs/2019APh...111..100A}
}

@ARTICLE{Kothes_2007A&A...468..993K,
       author = {{Kothes}, R. and {Dougherty}, S.~M.},
        title = "{The distance and neutral environment of the massive stellar cluster Westerlund 1}",
      journal = {\aap},
         year = 2007,
        month = jun,
       volume = {468},
       number = {3},
        pages = {993-1000},
          doi = {10.1051/0004-6361:20077309},
archivePrefix = {arXiv},
       eprint = {0704.3073},
 primaryClass = {astro-ph},
       adsurl = {https://ui.adsabs.harvard.edu/abs/2007A&A...468..993K}
}

@ARTICLE{Gagliardini_2024ApJ...969..161G,
       author = {{Gagliardini}, Silvia and {Langella}, Aurora and {Guetta}, Dafne and {Capone}, Antonio},
        title = "{Neutrino Fluxes from Different Classes of Galactic Sources}",
      journal = {\apj},
         year = 2024,
        month = jul,
       volume = {969},
       number = {2},
          eid = {161},
        pages = {161},
          doi = {10.3847/1538-4357/ad4960},
archivePrefix = {arXiv},
       eprint = {2403.05288},
 primaryClass = {astro-ph.HE},
       adsurl = {https://ui.adsabs.harvard.edu/abs/2024ApJ...969..161G}
}

@ARTICLE{Yadav_2017MNRAS.465.1720Y,
       author = {{Yadav}, Naveen and {Mukherjee}, Dipanjan and {Sharma}, Prateek and {Nath}, Biman B.},
        title = "{How multiple supernovae overlap to form superbubbles}",
      journal = {\mnras},
         year = 2017,
        month = feb,
       volume = {465},
       number = {2},
        pages = {1720-1740},
          doi = {10.1093/mnras/stw2522},
archivePrefix = {arXiv},
       eprint = {1603.00815},
 primaryClass = {astro-ph.HE},
       adsurl = {https://ui.adsabs.harvard.edu/abs/2017MNRAS.465.1720Y}
}

@ARTICLE{Blumenthal_1970RvMP...42..237B,
       author = {{Blumenthal}, George R. and {Gould}, Robert J.},
        title = "{Bremsstrahlung, Synchrotron Radiation, and Compton Scattering of High-Energy Electrons Traversing Dilute Gases}",
      journal = {Reviews of Modern Physics},
         year = 1970,
        month = jan,
       volume = {42},
       number = {2},
        pages = {237-271},
          doi = {10.1103/RevModPhys.42.237},
       adsurl = {https://ui.adsabs.harvard.edu/abs/1970RvMP...42..237B}
}

@ARTICLE{Aharonian_2010PhRvD..82d3002A,
       author = {{Aharonian}, F.~A. and {Kelner}, S.~R. and {Prosekin}, A. Yu.},
        title = "{Angular, spectral, and time distributions of highest energy protons and associated secondary gamma rays and neutrinos propagating through extragalactic magnetic and radiation fields}",
      journal = {\prd},
         year = 2010,
        month = aug,
       volume = {82},
       number = {4},
          eid = {043002},
        pages = {043002},
          doi = {10.1103/PhysRevD.82.043002},
archivePrefix = {arXiv},
       eprint = {1006.1045},
 primaryClass = {astro-ph.HE},
       adsurl = {https://ui.adsabs.harvard.edu/abs/2010PhRvD..82d3002A}
}

@BOOK{Draine_2011piim.book.....D,
       author = {{Draine}, Bruce T.},
        title = "{Physics of the Interstellar and Intergalactic Medium}",
         year = 2011,
       adsurl = {https://ui.adsabs.harvard.edu/abs/2011piim.book.....D}
}

\clearpage

\begin{appendix}

\section{Superbubble evolution}
\label{app:evolution}

The interaction between the collective wind of a compact star cluster and its ambient interstellar medium excavates an interstellar bubble. The gas swept by the forward shock is expected to form a thin shell around the bubble, with a total mass $M_{\mathrm{sh}} = 4\pi R_{\mathrm{b}}^3(t) \rho_0/3$, where $\rho_0$ is the mass density of the ambient medium and is assumed to be uniform here. The motion of the shell is governed by Newton's law, namely
\begin{equation}
    \frac{d}{dt} \left[ M_{\mathrm{sh}}(t) \dot{R}_{\mathrm{b}}(t) \right] = 4\pi R_{\mathrm{b}}^2(t) P,
    \label{eq:momentum}
\end{equation}
where $P$ is the shocked wind pressure.

The star cluster continuously injects kinetic energy into the bubble. The total energy in the bubble evolves as 
\begin{equation}
    \frac{d}{dt} \left[ \frac{4}{3} \pi R_{\mathrm{b}}^3(t) \frac{P}{\gamma - 1} \right] = L_{\mathrm{w}} - 4 \pi R_{\mathrm{b}}^2(t) P \dot{R}_{\mathrm{b}}(t) - L_{\mathrm{cool}},
    \label{eq:energy}
\end{equation}
where $L_{\mathrm{w}}=(1/2)\dot{M} v_{\mathrm{w}}^2$ is the wind kinetic luminosity, with $\dot{M}$ and $v_{\mathrm{w}}$ the mass loss rate of the star cluster and the wind velocity. $L_\mathrm{cool}$ represents the radiative cooling. Assuming $L_\mathrm{cool}=\zeta L_{\mathrm{w}}$, then $L_{\mathrm{w}}-L_\mathrm{cool}=(1-\zeta) L_{\mathrm{w}}=\xi_\mathrm{b} L_{\mathrm{w}}$, with $\xi_\mathrm{b}$ on the order of 0.1 \citep{Morlino_2021,Yadav_2017MNRAS.465.1720Y}. $P/(\gamma-1)$ is the gas energy density, with the adiabatic index $\gamma=5/3$ for the ideal monatomic gas. 

Equations \eqref{eq:momentum}--\eqref{eq:energy} can be solved by assuming $R_{\mathrm{b}} \propto t^{\alpha}$, which gives \citep{Castor_1975, Weaver_1977, Gupta_2018,Morlino_2021}
\begin{equation}
    R_{\mathrm{b}}(t) \simeq 0.76 ~ (\xi_{\mathrm{b}} L_{\mathrm{w}})^{1/5} \rho_{0}^{-1/5} t^{3/5}, \ \mathrm{and}
    \label{eq:Rb}
\end{equation}
\begin{equation}
    P(t) \simeq 0.16 ~ (\xi_{\mathrm{b}} L_{\mathrm{w}})^{2/5} \rho_{0}^{3/5} t^{-4/5}.
    \label{eq:pressure}
\end{equation}
On the other hand, the location of the termination shock can be calculated by balancing the pressure in the bubble and the ram pressure of the wind, which yields
\begin{equation}
    R_{\mathrm{s}}(t) = \left(\frac{\dot{M}v_{\mathrm{w}}}{4\pi P}\right)^{1/2} \simeq 0.7 ~ \dot{M}^{1/2} v_{\mathrm{w}}^{1/2} (\xi_{\mathrm{b}} L_{\mathrm{w}})^{-1/5} \rho_{0}^{-3/10} t^{2/5}.
\end{equation}
The average density and the temperature of the shocked wind is given by
\begin{equation}
    n_2 = \frac{3\dot M t}{4 \pi R_{\mathrm{b}}^3 m_\mathrm{H}} \simeq 3.25 \times 10^{23} ~ \dot{M} (\xi_{\mathrm{b}} L_{\mathrm{w}})^{-3/5} \rho_{0}^{3/5} t^{-4/5}, \ \mathrm{and}
    \label{eq:n2}
\end{equation}
\begin{equation}
    T_2 = \frac{P}{k_\mathrm{B} n_2} \simeq 3.57 \times 10^{-9}  ~ \dot{M}^{-1} \xi_{\mathrm{b}} L_{\mathrm{w}}.
\end{equation}

Thermal conduction between the cold shell and the hot bubble causes the evaporation of gas from the shell to the bubble. Assuming that the inward flow from the shell is approximated by a stationary plane-parallel flow and neglecting the radiative cooling in the transition zone, the flow at constant pressure can be described by \citep{Castor_1975,Menchiari_2023}
\begin{equation}
    \frac{\dot M_\mathrm{b}}{4\pi R_\mathrm{b}^2} \frac{dH}{dz} = \frac{d}{dz}\left[K(\tilde{T})\frac{d\tilde{T}}{dz}\right],
    \label{eq:thermal conduction}
\end{equation}
where $z=R_\mathrm{b}-r$. $H=5k_\mathrm{B}\tilde{T}/(2\mu)$ is the specific enthalpy, with $\mu=0.62m_\mathrm{H}$ the mean molecular weight. $K(\tilde{T})=C\tilde{T}^{5/2}$ is the thermal conductivity, with $C=1.2\times10^{-6}\,\mathrm{erg}\,\mathrm{cm}^{-1}\,\mathrm{s}^{-1}\,\mathrm{K}^{-7/2}$. $\dot M_\mathrm{b}$ is the mass gain rate of the bubble due to evaporation. The temperature in the bubble scales as $\tilde{T}(z)\propto z^{2/5}$, which can be rewritten as $\tilde{T}(r)=\tilde{T}_2 (1-r/R_\mathrm{b})^{2/5}$ for $R_\mathrm{b} \gg R_\mathrm{s}$, with $\tilde{T}_2$ the temperature immediately downstream \citep{Castor_1975}. The gas density follows $\tilde{n}(r) \tilde{T}(r)=\tilde{n}_2 \tilde{T}_2$ in the isobaric bubble. Integrating Equation \ref{eq:thermal conduction} from the shell ($z=0, \, \tilde{T}\approx0$) to the termination shock ($z\approx R_\mathrm{b}, \, \tilde{T}=\tilde{T}_2$) yields
\begin{equation}
    \dot M_\mathrm{b} = \frac{16\pi\mu}{25k_\mathrm{B}} K(\tilde{T}_2) R_\mathrm{b}.
    \label{eq:mass gain rate}
\end{equation}
After considering the evaporation of the shell, the density and temperature immediately downstream of the shock can be derived using Equation \ref{eq:Rb}, \ref{eq:pressure} and \ref{eq:mass gain rate}
\begin{equation}
    \tilde{n}_2 \simeq 2.8\times 10^{13} ~ (\xi_{\mathrm{b}} L_{\mathrm{w}})^{6/35} \rho_{0}^{19/35} t^{-22/35}, \ \mathrm{and}
\end{equation}
\begin{equation}
    \tilde{T}_2 \simeq 41.3 ~ (\xi_{\mathrm{b}} L_{\mathrm{w}})^{8/35} \rho_{0}^{2/35} t^{-6/35}.
    \label{eq:tildeT2}
\end{equation}
For $\rho_0 = 10 m_\mathrm{H} ~ \mathrm{cm}^{-3}$, $\dot{M} = 10^{-4} ~ M_{\odot} ~ \mathrm{yr}^{-1}$, $v_\mathrm{w} = 1000 ~ \mathrm{km} ~ \mathrm{s}^{-1}$, and $T_\mathrm{age} = 10 ~ \mathrm{Myr}$, we obtain $\tilde{n}_2 \approx 2.5\times10^{-2} \, \mathrm{cm}^{-3}$, $\tilde{T}_2 \approx 2.5 \times 10^6 \, \mathrm{K}$, and $n_2 \approx 3.5\times10^{-3} \, \mathrm{cm}^{-3}$, $T_2 \approx 1.8 \times 10^7 \, \mathrm{K}$. Therefore, the gas content in the bubble is primarily contributed by materials evaporated from the shell.

\section{Particle transport in the superbubble}
\label{app:transport}

Non-thermal particles propagate by diffusion and advection in the bubble excavated by the cluster wind. Their energy and spatial distributions can be obtained by solving the transport equation. Assuming spherical symmetry of the bubble, the transport equation is expressed as
\begin{equation}
    \frac{1}{r^2} \frac{\partial}{\partial r} \left[r^2 D \frac{\partial f}{\partial r}\right] - \frac{1}{r^2} \frac{\partial}{\partial r} \left[r^2\tilde{u} f\right] + \frac{1}{p^2} \frac{\partial}{\partial p} \left[p^2\dot{p} f\right] + Q = \frac{\partial f}{\partial t},
\end{equation}
where $f(r,p,t)$ is the particle distribution function in phase space. $Q(r,p,t)$ is the injection term, which can be expressed by $Q(r,p,t)=4 \pi R_{\mathrm{s}}^2 \eta_{\mathrm{inj}} n_1 u_1 [\delta(p - p_{\mathrm{inj}})/(4 \pi p^2)] [\delta(r - R_{\mathrm{s}})/(4 \pi r^2)]$. The energy loss term includes both the adiabatic energy loss/gain and the energy loss by interactions, namely $\dot{p}=\dot{p}_{\mathrm{ad}}+\dot{p}_{\mathrm{int}}$. $D(r,p,t)$ and $\tilde{u}(r,t)$ are the diffusion coefficient and the advection velocity, respectively.

Assuming the system has reached a steady state, the transport equation for relativistic particles can be written as
\begin{equation}
    \frac{1}{r^2} \frac{\partial}{\partial r} \left[r^2 D \frac{\partial N}{\partial r}\right] - \frac{1}{r^2} \frac{\partial}{\partial r} \left[r^2\tilde{u} N\right] + \frac{\partial}{\partial E} \left[\dot{E} N\right] + Q = 0,
\label{eq:transport2}
\end{equation}
where $N(r,E)$ is the distribution function in energy space, such that $N(r,E) dE=f(r,p) 4 \pi p^2 dp$. The injection term becomes $Q(r,E)=4 \pi R_{\mathrm{s}}^2 \eta_{\mathrm{inj}} n_1 u_1 \delta(E - E_{\mathrm{inj}}) \delta(r - R_{\mathrm{s}})/(4 \pi r^2)$, with $E_{\mathrm{inj}}$ the injection energy.
The diffusion coefficient can be written as 
\begin{equation}
    D(r,E) = 
    \begin{cases}
    D_1(r,E), & R_{\mathrm{sc}}<r \leq R_{\mathrm{s}} \\
    D_2(r,E), & R_{\mathrm{s}}<r \leq R_{\mathrm{b}}
    \end{cases}
\end{equation}
where $D_1$ and $D_2$ are the diffusion coefficients upstream and downstream of the shock respectively. $R_{\mathrm{sc}}$ is the size of the star cluster. For a detailed discussion of the diffusion coefficients, see Section \ref{sec:model}.
The advection velocity of non-thermal particles can be expressed as 
\begin{equation}
    \tilde u(r) = 
    \begin{cases}
    \tilde u_1, & R_{\mathrm{sc}}<r \leq R_{\mathrm{s}} \\
    \tilde u_2 \left(\frac{r}{R_{\mathrm{s}}}\right)^{-2}, & R_{\mathrm{s}}<r \leq R_{\mathrm{b}}
    \end{cases}
\end{equation}
where $\tilde u_1$ and $\tilde u_2$ are the effective velocities immediately upstream and downstream of the shock. The effective velocity is defined as the sum of the plasma speed $u(r)$ and the net speed of the turbulent waves that act as scattering centers, $\tilde u(r)=u(r)+\eta v_{\mathrm{A}}$, where $v_{\mathrm{A}}=B/\sqrt{4\pi \rho}$ is the Alfv\'en speed. Here we assume that waves propagate isotropically in the upstream ($\eta_1=0$), while anisotropy can exist in the downstream ($\eta_2\neq0$).

The total energy loss rate is given by the sum of adiabatic and interaction losses, $\dot{E} = \dot{E}_{\mathrm{ad}} + \dot{E}_{\mathrm{int}}$, where the adiabatic loss term is $\dot{E}_{\mathrm{ad}} = (\nabla \cdot \bm{\tilde{u}}) E / 3$.
For protons, energy losses due to interactions are dominated by proton-proton (pp) interactions, i.e., $\dot{E}_{\mathrm{int,p}} = \dot{E}_{\mathrm{pp}}$ \citep{Krakau_2015}. For electrons, we consider synchrotron emission, inverse Compton (IC) scattering, and bremsstrahlung. The total interaction loss rate is given by $\dot{E}_{\mathrm{int,e}} = \dot{E}_{\mathrm{syn}} + \dot{E}_{\mathrm{IC}} + \dot{E}_{\mathrm{brem}}$ \citep{Blumenthal_1970RvMP...42..237B}.

Boundary conditions are required to solve Equation \eqref{eq:transport2}. We assume that the particle flux is zero at the inner boundary, namely
\begin{equation}
    \left[u N-D \frac{\partial N}{\partial r}\right]_{R_{\mathrm{sc}}}=0, 
\end{equation}
and that particles can freely escape the system from the outer boundary: $N(R_{\mathrm{b}},E)=0$.

To obtain the particle spectrum at the shock, we first substitute the adiabatic energy loss rate into Equation \eqref{eq:transport2}, so that the transport equation becomes
\begin{equation}
\begin{aligned}
    \frac{\partial}{\partial r} \left[r^2 \left(\tilde{u}N - D \frac{\partial N}{\partial r}\right) \right] =
    &\frac{1}{3} \frac{d(r^2\tilde{u})}{dr} \frac{\partial}{\partial E}\left(EN\right) + r^2 \frac{\partial}{\partial E}\left(\dot{E}_{\mathrm{int}}N\right) \\
    &+ R_{\mathrm{s}}^2 \eta_{\mathrm{inj}} n_1 u_1 \delta(E - E_{\mathrm{inj}}) \delta(r - R_{\mathrm{s}}).
\end{aligned}
\label{eq:transport3}
\end{equation}
We then integrate Equation \eqref{eq:transport3} from $R_{\mathrm{s}}^{-}$ to $R_{\mathrm{s}}^{+}$, which yields
\begin{equation}
\begin{aligned}
    -D_2 \frac{\partial N}{\partial r} \bigg|_2 + D_1 \frac{\partial N}{\partial r} \bigg|_1 = 
    &\frac{1}{3} (\tilde{u}_2 - \tilde{u}_1) E \frac{dN_0}{dE} - \frac{2}{3} (\tilde{u}_2 - \tilde{u}_1) N_0 \\ 
    &+ \eta_{\mathrm{inj}} n_1 u_1 \delta(E - E_{\mathrm{inj}}).
\end{aligned}
\label{eq:transport4}
\end{equation}
By introducing
\begin{equation}
    \mathcal{D}_{1,2}(E) = \frac{D_{1,2} \left. \frac{\partial N}{\partial r} \right|_{1,2}}{\tilde{u}_{1,2} N_0(E)},
\end{equation}
Equation \eqref{eq:transport4} becomes
\begin{equation}
\begin{aligned}
    &\frac{dN_0}{dE} + \left[ \frac{3\mathcal{D}_2(E)}{1-\mathcal{R}} - \frac{3\mathcal{R}(\mathcal{D}_1(E) - 1)}{1-\mathcal{R}} + \frac{\mathcal{R} + 2}{\mathcal{R} - 1} \right] \frac{N_0}{E} \\
    &= \frac{3u_1}{\tilde u_1-\tilde u_2} \frac{\eta_{\mathrm{inj}} n_1 \delta(E - E_{\mathrm{inj}})}{E},
\end{aligned}
\label{eq:transport5}
\end{equation}
where $\mathcal{R}=\tilde u_1 / \tilde u_2$ is the effective compression factor. The solution for $N_0(E)$ at $E>E_{\mathrm{inj}}$ can be expressed as 
\begin{equation}
   N_0(E) = K \left(\frac{E}{E_{\mathrm{inj}}}\right)^{-\frac{\mathcal{R}+2}{\mathcal{R}-1}} \times \exp \left\{ \int_{E_{\mathrm{inj}}}^{E} \frac{dE'}{E'} \left[\frac{3\mathcal{R}(\mathcal{D}_{1}-1)}{1-\mathcal{R}} - \frac{3\mathcal{D}_2}{1-\mathcal{R}}\right] \right\},
\label{eq:solution}
\end{equation}
with 
\begin{equation}
    K= \frac{3u_1}{\tilde u_1-\tilde u_2} \frac{\eta_{\mathrm{inj}} n_1}{E_{\mathrm{inj}}}.
\end{equation}
The spectrum at the shock is normalized by an efficiency of conversion ($\xi_{\mathrm{CR}}$) of the wind ram pressure to the pressure of non-thermal particles.
Equation \eqref{eq:solution} is an implicit solution for the particle spectrum at the shock, since $\mathcal{D}_{1,2}(E)$ depends on the diffusive fluxes immediately upstream and downstream of the shock, which are unknown a priori. To solve for $N(r,E)$, we adopt the numerical iterative algorithm developed by \cite{Blasi_2023}. 

\section{Non-thermal emission from the superbubble}
\label{app:emission}

The gamma-ray emissivity from pp interactions can be calculated as (in units of $\mathrm{cm}^{-3} ~ \mathrm{s}^{-1} ~ \mathrm{TeV}^{-1}$)
\begin{equation}
    J_{\mathrm{pp}}(r, E_{\gamma}) = cn_\mathrm{b} \int_{E_\gamma} N_\mathrm{p}(r,E) dE ~ \sigma_{\mathrm{inel}}(E) \frac{F_\gamma \left( \frac{E_\gamma}{E}, E \right)}{E}, 
\end{equation}
where $F_\gamma \left( E_\gamma / E, E \right)$ is the gamma-ray spectrum per collision, and $\sigma_{\mathrm{inel}}(E)$ is the inelastic cross section of pp interactions \citep{Kelner_2006,Kafexhiu_2014}. $n_\mathrm{b}$ is the number density of target gas. The production of secondary neutrinos and electrons can also be calculated in a similar way.

For electrons, the synchrotron emissivity is given by \citep{Blumenthal_1970RvMP...42..237B, Aharonian_2010PhRvD..82d3002A}
\begin{equation}
    J_{\mathrm{syn}}(r, E_{\gamma}) = \frac{\sqrt{3} e^3 B}{h m_\mathrm{e} c^2} \frac{1}{E_\gamma} \int N_\mathrm{e}(r,E) dE ~ F \left( \frac{E_\gamma}{E_\mathrm{c}} \right), 
\end{equation}
where $F(E_\gamma / E_\mathrm{c})$ is the synchrotron function, with $E_\mathrm{c}=3ehB\gamma^2/(4\pi m_\mathrm{e}c)$ the characteristic energy.

The IC emissivity takes the form
\begin{equation}
\begin{aligned}
    J_{\mathrm{IC}}(r, E_{\gamma}) 
    & = \frac{2\pi r_0^2 c}{\gamma^2} \iint N_\mathrm{e}(r,E) dE ~ \frac{n(\epsilon)d\epsilon}{\epsilon} \\
    & \times \left[ 2q\ln{q} + (1+2q)(1-q) + \frac{(\Gamma_\epsilon q)^2 (1-q)}{2(1+\Gamma_\epsilon q)} \right],
\end{aligned}
\end{equation}
where $\Gamma_\mathrm{e}=4\epsilon \gamma/(m_\mathrm{e} c^2)$ and $q=E_\gamma/[\Gamma_\mathrm{e}(E-E_\gamma)]$ \citep{Blumenthal_1970RvMP...42..237B}. $n(\epsilon)$ is the differential number density of background photons. $r_0$ is the classical electron radius.

Similarly, the bremsstrahlung emissivity is written as 
\begin{equation}
\begin{aligned}
    J_{\mathrm{brem}}(r, E_{\gamma}) 
    & = c n_\mathrm{b} \alpha r_0^2 \frac{1}{E_\gamma} \int N_\mathrm{e}(r,E) dE  \\
    & \times \left[ (1+(1-\frac{E_\gamma}{E})^2) \phi_1 - \frac{2}{3} (1-\frac{E_\gamma}{E}) \phi_2 \right],
\end{aligned}
\end{equation}
where $\phi_1=\phi_2=8[\ln{(2E(E-E_\gamma)/(m_\mathrm{e} c^2 E_\gamma))} - 1/2]$ for fully ionized hydrogen gas \citep{Blumenthal_1970RvMP...42..237B}. $\alpha$ is the fine-structure constant.

The radiative intensity is then obtained by integrating the emissivity along the line of sight
\begin{equation}
    I(\theta, E_{\gamma}) = \frac{1}{4\pi} \int_{l_{\mathrm{-}}}^{l_{\mathrm{+}}} J\left[r(\theta,l), E_{\gamma}\right] dl,
\end{equation}
where $\theta$ is the angle between the line of sight and the direction toward the center of the bubble, and $l$ is the distance from the observer to a point along the line of sight, with $l_{\mathrm{\pm}}=d \cos{\theta} \pm \sqrt{R_{\mathrm{b}}^2-d^2 \sin^2{\theta}}$ and $d$ the distance to the source center.  The integration is divided by $4\pi$ to account for the assumption of isotropic emission.
The total flux from the bubble is calculated as
\begin{equation}
    \Phi(E_{\gamma}) = 2\pi \int I(\theta, E_{\gamma}) \sin{\theta} d\theta,
\end{equation}
in units of $\mathrm{cm}^{-2} ~ \mathrm{s}^{-1} ~ \mathrm{TeV}^{-1}$.

\end{appendix}

\end{document}